\documentstyle[epsf]{article}

\newcommand{\newsection}{ \setcounter{equation}{0} \section}
\newcommand{\beq}{\begin{equation}} \newcommand{\eeq}{\end{equation}}
\newcommand{\bea}{\begin{eqnarray}} \newcommand{\eea}{\end{eqnarray}}

  \newcommand
{\Romannumeral}[1]{\uppercase\expandafter{\romannumeral#1}}

\newcommand{\be}{\begin{enumerate}} \newcommand{\ee}{\end{enumerate}}
\newcommand{\bi}{\begin{itemize}} \newcommand{\ei}{\end{itemize}}
\newcommand{\ba}{\begin{array}} \newcommand{\ea}{\end{array}}
\newcommand{\bc}{\begin{center}} \newcommand{\ec}{\end{center}}
\newcommand{\bt}{\begin{tabular}} \newcommand{\et}{\end{tabular}}

\def\lsim{\mathrel{\rlap{\lower4pt\hbox{\hskip1pt$\sim$}}
    \raise1pt\hbox{$<$}}}           
\def\gsim{\mathrel{\rlap{\lower4pt\hbox{\hskip1pt$\sim$}}
    \raise1pt\hbox{$>$}}}           

\newcommand{\half}{\textstyle {1\over2} \displaystyle}    
\newcommand{\Dslash}{{\hbox{D}\kern-0.6em\raise0.15ex\hbox{/}}} 

\begin{document}

\setlength{\oddsidemargin}{0cm} \setlength{\baselineskip}{7mm}

\input epsf




\begin{normalsize}\begin{flushright}
    UCI-99-20 \\
    September 1999 \\
\end{flushright}\end{normalsize}

\begin{center}
  
\vspace{50pt}
  
{\Large \bf On the Gravitational Scaling Dimensions }

\vspace{40pt}
  
{\sl Herbert W. Hamber}
$^{}$\footnote{e-mail address : hhamber@uci.edu;
http://aeneas.ps.uci.edu/hamber}
\\

\vspace{20pt}

Department of Physics and Astronomy \\
University of California \\
Irvine, CA 92171 \\

\end{center}

\vspace{40pt}

\begin{center} {\bf ABSTRACT } \end{center}
\vspace{12pt}
\noindent

A model for quantized gravitation based on the simplicial lattice
discretization is studied in detail using a comprehensive finite
size scaling analysis combined with renormalization group methods.
The results are consistent with a value for the universal 
critical exponent for gravitation $\nu=1/3$, and suggest a simple
relationship between Newton's constant, the gravitational
correlation length and the observable average space-time curvature.
Some perhaps testable phenomenological implications of these results
are discussed.
To achieve a high numerical accuracy in the evaluation of
the lattice path integral a dedicated parallel machine was assembled.

\vspace{24pt}

\begin{center} {\it (Submitted to the Physical Review D)} \end{center}

\vfill

\newpage

\vskip 10pt
\newsection{Introduction}
\hspace*{\parindent}

One of the outstanding problems in theoretical physics is
a determination of the quantum-mechanical properties
of Einstein's relativistic theory of Gravitation.
Approaches based on linearized perturbation methods have had moderate
success so far,
as the underlying theory is known not to be perturbatively renormalizable
\cite{fey,hooft}.
Due to the complexity of even such approximate calculations
a fundamental coupling of the theory, the bare cosmological constant
term, is usually set to zero thus further restricting the potential physical
relevance of the results.
In addition gravitational fields are themselves the source for
gravitation already at the classical level,
which leads to the problem of a intrinsically non-linear theory, where
perturbative results are possibly of doubtful validity for sufficiently
strong effective couplings.
This is especially true in the quantum domain, where large fluctuations
in the gravitational field appear at short distances.
In general nonperturbative effects can give rise to novel
behavior in a quantum field theory, and in
particular to the emergence of non-trivial fixed points of the renormalization
group (a phase transition in statistical mechanics language).
It has been realized for some time that in general the universal 
low and high energy behavior of field theories is almost completely
determined by the fixed point structure of the renormalization group
trajectories \cite{wilson}.

The situation described above bears some resemblance to the theory
of strong interactions, Quantum Chromodynamics. Non-linear
effects are known here to play an important role, and end up restricting
the validity of perturbative calculations to the high energy, short distance
regime, where the effective gauge coupling can be considered weak due
to asymptotic freedom \cite{gross}.
For low energy properties Wilson's discrete lattice formulation, combined with
the renormalization group and computer simulations, has provided so far
the only convincing evidence for quark confinement and chiral
symmetry breaking, two phenomena which are invisible to any order in the
weak coupling, perturbative expansion.

A discrete lattice formulation can
be applied to the problem of quantizing gravitation.
Instead of continuous metric fields, one deals with gravitational
degrees of freedom which live only on discrete space-time points and
interact locally with each other.
In Regge's simplicial formulation of gravity \cite{regge} one approximates
the functional integration over continuous metrics by a discretized
sum over piecewise linear simplicial geometries \cite{rowi,lesh,hartle,hw84}.
In such a model, the role of the continuum metric is played
by the edge lengths of the simplices, while curvature is
described by a set of deficit angles, which can be computed via
known formulae as functions of the given edge lengths.
The simplicial lattice
formulation of gravity is locally gauge invariant \cite{gauge} and can be
shown to contain perturbative gravitons in the lattice weak field expansion
\cite{rowi}, making it an attractive and faithful lattice regularization
of the continuum theory.

The discretized theory is restricted to a finite set of dynemical
variables, once a set of suitable boundary conditions are imposed
such as periodic or with some assigned boundary manifold.
In the end the original continuum theory of gravity is to be recovered
as the space-time volume is made large and the fundamental lattice
spacing  of the discrete theory is sent to zero. Possibly without
having to rely, at least in principle, on any further approximation
to the original continuum theory.

Quantum fluctuations in the underlying geometry are represented
in the discrete theory
by fluctuations in the edge lengths, which can be modeled by
a well-defined, and numerically exact, stochastic process.
In analogy with other field theory models studied by computer,
calculations are usually performed in the Euclidean
imaginary time framework, which is the only formulation amenable
to a controlled numerical study, at least for the immediate foreseeable future.
The Monte-Carlo method, based on the concept of importance sampling,
is well suited for evaluating the discrete path integral for
gravity and for computing the required averages and correlation
functions.
By a careful and systematic analysis of the lattice results, the
critical exponents can be extracted, and the scaling properties of
invariant correlation functions determined from first principles.

Studies on small lattices suggest a rich scenario
for the ground state of quantum gravity \cite{lesh,hw84,monte,phases}.
The present evidence indicates that simplicial quantum gravity
in four dimensions exhibits a phase transition (in the bare coupling $G$)
between {\it two phases}:
a strong coupling phase, in which the geometry is smooth at large
scales and quantum fluctuations in the gravitational field eventually
average out and are bounded;
and a weak coupling phase, in which the geometry is degenerate
and space-time collapses into a lower-dimensional manifold, bearing
some physical resemblance to a branched polymer.
Only the smooth, small negative curvature and thus anti-DeSitter-like
phase appears to be physically acceptable.
Phrased in different terms, the two phases of quantized gravity found
in ~\cite{phases}, can loosely be described as having in one phase
(with bare coupling $G<G_c$, the rough branched polymer-like phase)
\beq
\langle g_{\mu\nu} \rangle \; = \; 0 \;\; ,
\eeq
while in the other (with bare coupling $G>G_c$, the smooth phase),
\beq
\langle g_{\mu\nu} \rangle \; \approx \; c \; \eta_{\mu\nu} \;\; ,
\eeq
with a vanishingly small negative average curvature in the vicinity
of the critical point at $G_c$.
The existence of a phase transition at finite
coupling $G$, usually associated in quantum field theory with the appearance
of an ultraviolet fixed point of the renormalization group, implies
in principle non-trivial, calculable non-perturbative scaling properties
for correlations and effective coupling constants, and in particular
in the case at hand for Newton's gravitational constant.
Since only the smooth phase with $G>G_c$ has acceptable
physical properties, one would conclude on the basis of fairly general
renormalization group arguments that at least in this lattice model the 
gravitational coupling can only {\it increase} with distance.
Furthermore, the rise of the gravitational coupling in the infrared
region rules out the applicability of perturbation theory
to the low energy domain, to the same extent that such an
approach is deemed to be inapplicable to study the low-energy properties
of asymptotically free gauge theories.

It is a remarkable property of quantum field theories that
a wide variety of physical properties can be determined from a relatively
small set of universal quantities \cite{kawi}. Namely the
universal leading critical exponents, computed in the vicinity of some
fixed point (or fixed line) of the renormalization group equations.
In the lattice theory the presence of a fixed point or phase transition is
often inferred from the appearance of non-analytic terms in invariant local
averages, such as for example the average curvature
\beq
<l^2> { < \int d^4 x \, \sqrt{ g } \, R(x) >
\over < \int d^4 x \, \sqrt{ g } > }
\; \equiv \; {\cal R} (k)
\mathrel{\mathop\sim_{ k \rightarrow k_c}}
- A_{\cal R} \; ( k_c - k ) ^{ 4 \nu - 1 } \;\; ,
\eeq
where $k=1/8 \pi G$. From such averages one can determine the value for
$\nu$, the correlation length exponent,
\beq
\xi (k) \; \mathrel{\mathop\sim_{ k \rightarrow k_c}} \; A_\xi \;
( k_c - k ) ^{ -\nu } \;\; .
\eeq
An equivalent result, relating the quantum expectation value
of the curvature to the physical correlation length $\xi$ , is
\beq
{\cal R} ( \xi ) \; \mathrel{\mathop\sim_{ k \rightarrow k_c}} \;
\xi^{ 1 / \nu - 4 }
\;\; .
\eeq
Matching of dimensionalities in these equations is restored by
supplying appropriate powers of the ultraviolet cutoff, the Planck
length $l_P=\sqrt{G}$.
The exponent $\nu$ is known to be related to the derivative of the beta
function for $G$ in the vicinity of the ultraviolet fixed point,
\beq
\beta ' (G_c) \, = \, - 1/ \nu \;\; .
\eeq
In addition, the correlation length $\xi$ itself determines the
long-distance decay of the connected, invariant two-point correlations at
fixed geodesic distance $d$.
For the curvature correlation one has for distances much larger compared
to the correlation length
\beq
< \sqrt{g} \; R(x) \; \sqrt{g} \; R(y) \; \delta ( | x - y | -d ) >_c \;
\mathrel{\mathop\sim_{d \; \gg \; \xi }} \;\;
d^{- \sigma } \; e^{-d / \xi } \;\; ,
\eeq
while for shorter distances one expects a slower power law decay
\beq
< \sqrt{g} \; R(x) \; \sqrt{g} \; R(y) \; \delta ( | x - y | -d ) >_c \;
\mathrel{\mathop\sim_{d \; \ll \; \xi }} \;\; 
{1 \over d^{\; 2 \; (4-1/ \nu)} } \;\; .
\eeq
The possibility of non-trivial scaling dimensions in the theory
of gravitation is not new and was pointed out some time ago in a
series of interesting papers \cite{fubini}.
Moreover it is easy to see that the scale dependence of the
effective Newton constant is given by
\beq
G(r) \; = \; G(0) \left [ \, 1 \, + \, c \, ( r / \xi )^{1 / \nu} \, 
+ \, O (( r / \xi )^{2 / \nu} ) \, \right ] \;\; ,
\label{eq:grun}
\eeq
with $c$ a calculable numerical constant.
In this last expression the momentum scale $\xi^{-1}$ plays a role
similar to the
scaling violation parameter $\Lambda_{\overline{MS}}$ of QCD.
It seems natural, although paradoxical at first, to associate $\xi$ with
some macroscopic cosmological length scale, such as the Hubble distance
$c H_0^{-1}$, with the lack of screening of gravitational
interactions ultimately accounting for such an unusual interpretation
\cite{phases,det}.
Of course an increase of the gravitational coupling at large distances
signals a likely breakdown of perturbation theory for computing low
energy properties of gravity.

It should be clear, even from this brief discussion, that the
critical exponents by themselves already provide a significant
amount of useful information about the continuum theory.
In reality, the complexity of the lattice interactions and the practical
need to sample
many statistically independent field configurations contributing to the
path integral, which is necessary for correctly incorporating
into the model the effects of quantum-mechanical
fluctuations, leads to the requirement of powerful computational
resources.
The results presented in this paper were obtained using
a dedicated custom-built 20-GFlop 64-processor parallel
computer, described in detail in \cite{aeneas}.

Finally one should mention that recently there has been a significant
resurgence of interest in the classical applications of the Regge
formulation to gravity.
A description of the methods as applied to several aspects of the
initial value problem in General Relativity can be found
in the recent references in \cite{miller}.
For a related approach to lattice gravity
based on dynamical triangulations see also \cite{smit}.

A brief outline of the paper is as follows.
Section 2.\ contains a discussion of general and finite size scaling
and related issues as they apply to the lattice theory of gravity.
Section 3.\ touches on the issue of the unboundedness of
the Euclidean gravitational action.
Section 4.\ defines local curvature averages and their fluctuations, while
Section 5.\ introduces a set of exact sum rules for averages which
follow from the scaling properties of the partition function.
Section 6.\ defines a set of invariant correlations and discusses
how they relate to the local fluctuations defined previously.
Section 7.\ includes a general discussion of the expected properties of
the theory in the presence of an ultraviolet fixed point, including
expectations based on the analytical $2+\epsilon$ expansion.
In Section 8.\ the numerical results are presented.
Section 9.\ contains a discussion of the possible future physical
relevance of the results, while Section 10.\ contains the conclusions.

\vskip 30pt
\section{Finite Size Scaling}
\hspace*{\parindent}

One of the most important quantities used in establishing the continuum
limit of a lattice field theory are the critical exponents.
Reliable estimates for the exponents in a lattice field theory
require a comprehensive finite-size analysis, a procedure by which
accurate values for the critical exponents are obtained by taking into
account the linear size dependence of the result computed in a finite
volume $V$.
One starts from the general Euclidean Action (or statistical mechanics
Hamiltonian)
\beq
H \; = \; \sum_i \; g_i \; {\cal O}_i
\label{eq:ham}
\eeq
with $g_i$ the coupling associated with the operator ${\cal O}_i$.
In the gravitational case the couplings would correspond to the
bare cosmological constant, the Newtonian gravitational constant
and the higher derivative coupling.
Close to a renormalization group fixed point denoted by $\{ g^*_i \}$
one chooses the ${\cal O}_i$'s to be eigenvectors of the linearized
renormalization group transformation, such that
\beq
g_i - g^*_i \; \rightarrow \; b^{y_i} ( g_i - g^*_i ) \;\;\; ,
\eeq
where $b$ is the scale factor of the transformation.
In the simplest statistical mechanics systems, such as a ferromagnet
in the absence of an external magnetic field, one has ${\cal O} \sim H $ as
the only relevant operator (in the sense that $y>0$),
and $g \sim t=T-T_c$.
As will be discussed below, in the gravitational case the role
of $T$ is played by the bare gravitational coupling $G$.
Additional operators appearing in the action
are classified as marginal
$(y=0)$ or irrelevant. The relevance of the energy operator reflects the
fact that close to the critical point $t$ is the only parameter that
needs to be tuned to achieve criticality, synonymous with long range
correlations.
Universality of critical behavior then accounts for the fact that many
diverse physical systems exhibit the same scaling behavior in the vicinity
of the critical point, as a consequence of a divergent correlation length
\cite{kawi}.

In practice the renormalization group approach is brought in via a
slightly different route, involving a change in the overall
linear size of the system.
The usual starting point for the derivation of the scaling properties
of the theory is the Renormalization Group (RG) behavior of the free
energy $F\, = \, - \log Z / V$
\beq
F(t, \{ u_j \} ) \; = \; F_{reg} (t, \{ u_j \} ) \; + \;
b^{-d} \; F_{sing} ( b^{y_t} t, \{ b^{y_j} u_j \} ) \;\; ,
\label{eq:fsing}
\eeq
where $F_{sing}$ is the singular, non-analytic part of the free
energy, and $F_{reg}$ is the regular part. $b$ is the block size in
the RG transformation, while $y_t$ and $y_j (j \ge 2)$
are the relevant eigenvalues of the RG transformation
(for more details see the review \cite{barber}). 
One denotes here by $y_t > 0 $ the relevant eigenvalue, while the remaining
eigenvalues $y_j \le 0 $ are associated with either marginal or irrelevant
operators.
Usually $y_t^{-1}$ is called $\nu$, while the next subleading exponent
$y_2$ is denoted $-\omega$.

The correlation length $\xi$ determines the asymptotic decay of correlations,
in the sense that one expects for example for the two-point function
at large distances
\beq
< {\cal O} (x) {\cal O} (y) > \;
\mathrel{\mathop\sim_{ |x-y| \; \gg \; \xi }} \; e^{-|x-y| / \xi } \;\;\; .
\label{eq:xi}
\eeq
The scaling equation for the correlation length itself
\beq
\xi(t) \; = \; b \; \xi \left ( b^{y_t} t \right )
\eeq
implies for $b = t^{-1/y_t}$ that $ \xi \sim t^{- \nu}$ with a
correlation length exponent
\beq
\nu = 1/y_t \;\; .
\label{eq:nu}
\eeq
Derivatives of the free energy $F$ with respect to $t$ then determine,
after setting the scale factor $b=t^{-1/y_t}$,
the scaling properties of physical observables, including corrections
to scaling \cite{wegner}. Thus for example, the second derivative
of the free energy with respect to $t$ yields the specific heat
exponent $\alpha = 2 - d/ y_t = 2-d \nu$,
\beq
{ \partial^2 \over \partial t^2 } \;  F(t, \{ u_j \} )
\; \sim \; t^{-(2-d \nu)} \;\;\;\; .
\label{eq:cv_sing}
\eeq
In the gravitational case one identifies the scaling field $t$
with $k_c - k$, where $k=1/16 \pi G$ involves the bare Newton's constant.
The appearance of singularities in physical averages, obtained from
appropriate derivatives of $F$, is rooted in the fact that
close to the critical point at $t=0$ the correlation length diverges.

The above results can be extended to the case of a finite lattice
of volume $V$ and linear dimension $L=V^{1/d}$.
The volume-dependent free energy is then written as
\beq
F(t, \{ u_j \}, L^{-1} ) \; = \; F_{reg} (t, \{ u_j \} ) \; + \;
b^{-d} \; F_{sing} ( b^{y_t} t, \{ b^{y_j} u_j \}, b/L ) \;\;\; .
\label{eq:fssf}
\eeq
For $b=L$ (a lattice consisting of only one point) one obtains the 
Finite Size Scaling (FSS) form of the free energy
(for a detailed presentation of this procedure see \cite{blote};
see also \cite{qftfss,iz} for a field-theoretic justification).
After taking derivatives
with respect to the fields $t$ and $ \{ u_j \} $, the FSS
scaling form for physical observables follows.
For a quantity $O$ diverging like $ t^{-x_{O}}$ in the infinite
volume limit one has
\beq
O(L,t) \; = \; L^{x_O / \nu} \; \left [ \tilde{f_O}
\left ( { L \over \xi( \infty ,t) } \right ) \; + \; 
{\cal O} (\xi^{-\omega}, L^{-\omega}) \right ] \;\;\; ,
\eeq
with $\tilde{f_O}$ a smooth scaling function, and 
$\xi(\infty,t)$ the infinite volume correlation length.
For sufficiently large
volumes the correction to scaling term involving $\omega$ can be neglected,
but in general one needs to be aware of their presence if either
the volumes are not large enough or if the corrections are large due to a
large amplitude or small exponent.
Some properties of the scaling function $\tilde{f_O}(y)$ can be
deduced on general grounds: it is expected to show a peak if the
finite volume value for $O$ is peaked, it
is analytic at $x=0$ since no singularity can develop in a finite
volume, and $\tilde{f_O}(y) \sim \tilde y^{-x_O} $ for large $y$ for a
quantity $O$ which diverges as $t^{-x_O}$ in the infinite volume limit.

The last expression is useful when the infinite-volume correlation length
is known. But since close to the critical point $ \xi \sim t^{- \nu}$, 
one can deduce the equivalent scaling from
\beq
O(L,t) \; = \; L^{x_O / \nu} \; \left [ \tilde{f_O}
\left ( L \; t^{\nu} \right ) \; + \; 
{\cal O} ( L^{-\omega}) \right ] \;\;\; ,
\label{eq:fsso}
\eeq
which relies on a knowledge of $t$, and thus of the critical point, instead.
For a state of the art application of the above methods to the 3-d Ising
model see \cite{ising}.

The previous discussion applies to continuous, second order phase transitions.
First order phase transitions are driven by instabilities, and are in
general not 
governed by any renormalization group fixed point. The underlying reason
is that the correlation length does not diverge at the transition point,
and thus the system never becomes scale invariant.
Exponents for continuous, second order phase transition in general obey the
rigorous bound
\beq
y_t < d  \;\;\;\;  or \;\;\;\; \nu > 1/d  \;\; .
\eeq
First order phase transition in renormalization group theory,
on the other hand, can be associated with the somewhat
pathological case $\nu=1/d$, for which the first derivative
of the free energy develops a step-function singularity.
In a renormalization group framework the corresponding pseudo-critical
point is denoted as a discontinuity fixed point \cite{nn}.

In the simplest case, a first order transition develops as the system
tunnels between two neighboring minima of the free energy.
In the metastable branch the free energy acquires a complex part
with an essential singularity in the coupling located at the first order
transition point \cite{pbook,langer}.
As a consequence, such a singularity is not generally visible from the stable
branch, in the sense that a power series expansion in the
temperature is unaffected by such a singularity.
Indeed in the infinite volume limit the singularity associated
with a first order transition at $T_c$ becomes infinitely sharp,
like a $\delta-$ or $\theta$-function type singularity.
The singularity in the free energy at the endpoint of the metastable branch
(at say $T^*$) then cannot be explored directly, it has to reached by an
analytic continuation from the stable side of the free energy branch.

\vskip 30pt
\section{Unboundedness of the Euclidean Theory}
\hspace*{\parindent}

Perturbation theory on a lattice and in the continuum suggests the
presence of an instability in the Euclidean formulation for
sufficiently smooth manifold.
It is also known that the above instability
is associated with the appearance of a wrong sign for the conformal mode.
On the lattice the instability seems to persist close to the critical
point \cite{phases}, which suggests that the continuum limit has to be
reached by some sort of analytic continuation from the stable phase towards
the critical point, naturally defined as the point in coupling constant
space where the correlation length diverges.

In the weak-field expansion \cite{veltman} the Einstein-Hilbert
action contains both spin two (graviton) and spin zero (conformal
mode) contributions.
In the continuum one can by a judicious choice of invariant correlation
functions isolate physical properties of the graviton from the conformal mode.
A similar result holds on the lattice,
as can be seen by expanding the Regge action
about a regular lattice and using the fact that the lattice and continuum
actions are equivalent for sufficiently smooth manifolds \cite{rowi,hw3d}.
In general, after expanding the metric around flat space
(which requires $\lambda=0$),
\beq
g_{\mu\nu} \; = \; \eta_{\mu\nu} + \sqrt{16 \pi G} \; h_{\mu\nu} \;\;\;  ,
\eeq
one can cast the lowest order quadratic contribution to the
action in the form
\beq
I_{E} [ h_{\mu\nu} ] = \half \int d^4 x  \;
h_{\mu\nu} V_{\mu\nu\lambda\sigma} h_{\lambda\sigma}  \;\;\; ,
\eeq
where $V$ is a matrix which can be expressed in terms of spin projection
operators. In momentum space it can be written as
\beq
V = \bigl [ P^{(2)} - 2 P^{(0)} \bigr ] \;  p^2 \;\;\; ,
\eeq
where $P^{(2)}$ and $P^{(0)}$ are spin two and spin zero
projection operators introduced in \cite{vnh}.
Physically, the two terms correspond to the propagation of the graviton
and of the conformal mode, respectively, with the latter one appearing
with the `wrong' sign.
In the `Landau' gauge, with a gauge fixing term 
$\alpha^{-1} (\partial_\mu h^{\mu\nu})^2 $ and $\alpha=0$,
one obtains for the graviton propagator in momentum space
\beq
G_{\mu\nu\lambda\sigma} (p) \; = \;
{ P^{(2)}_{\mu\nu\lambda\sigma} \over p^2 } -
{ \half P^{(0)}_{\mu\nu\lambda\sigma} \over p^2 } \;\;\; .
\label{eq:gravprop}
\eeq
The unboundedness of the Euclidean gravitational action shows up
clearly in the weak field expansion, with the spin zero mode acquiring a
propagator term with the wrong sign.
$^{}$\footnote{It should be noted that such an instability is not peculiar
to gravitation.
Indeed the Euclidean path integral for the one-dimensional Coulomb potential,
an otherwise completely well behaved quantum mechanical system,
already exhibits such an instability. It would be premature to 
conclude from such a result that the problem is physically ill-posed.}
It has been argued that in weak field perturbation
theory and in order to avoid the unboundedness problem one should perform
the functional integral over metrics by distorting the integration contour
so as to include complex conformal factors \cite{haw}.
One drawback of this prescription is that it only appears applicable
within the framework of perturbation theory.
For a recent review of the Euclidean instability problem see \cite{moda99}.

In the presence of a cosmological constant, things are further
complicated by the fact that since flat space is no longer
a solution of the classical equations of motion, and the above
expansion for the metric looses part of its meaning due to the presence
of the tadpole term. But after shifting to the correct 0-th order
solution, a similar result is obtained. One can further modify the
action to include additional invariant terms, but things do not get
any better.
In the presence of higher derivative terms in the gravitational
action, the above result is
modified by terms $O(p^4)$, and becomes \cite{hdqg}
\beq
G_{\mu\nu\lambda\sigma} (p) \; = \;
{ P^{(2)}_{\mu\nu\lambda\sigma} \over p^2 + {2a \over k} p^4 } +
{ \half P^{(0)}_{\mu\nu\lambda\sigma} \over -p^2 + {a \over k} p^4 } \;\;\; .
\label{eq:gravprop2}
\eeq
The $p^4$ terms improve the ultraviolet behavior of the theory,
but do not remove the unboundedness problem, which re-appears
for sufficiently small $p^2$, in the low momentum or long-distance
limit.
Moreover, the resulting theory is most likely not unitary unless
the coupling $a$ is vanishingly small. The lack of positivity
of physical correlations for $a>0$ can be seen explicitly 
even in a non-perturbative treatment \cite{corr}, and makes such
a modified theory of gravitation in the end somewhat unattractive.

\vskip 30pt
\section{Local Averages and Fluctuations}
\hspace*{\parindent}

In the following the relevant definitions for
gravitational averages and correlations on the lattice will be
briefly recalled, in a form which will be used in later sections.
The starting point for a non-perturbative study of quantum gravity
is a suitable definition of the discrete Feynman path integral.
In the simplicial lattice approach one starts from the discretized
Euclidean path integral for pure gravity,
with the squared edge lengths taken as fundamental variables,
\beq
Z_L \; = \; \int_0^\infty \; \prod_s \; \left ( V_d (s) \right )^{\sigma} \;
\prod_{ ij } \, dl_{ij}^2 \; \Theta [l_{ij}^2]  \; 
\exp \left \{ 
- \sum_h \, \Bigl ( \lambda \, V_h - k \, \delta_h A_h 
+ a \, \delta_h^2 A_h^2 / V_h  + \cdots \Bigr ) \right \}  \;\; .
\label{eq:zlatt} 
\eeq
The above expression represents a lattice discretization of the
continuum Euclidean path integral for pure quantum gravity
\beq
Z_C \; = \; \int \prod_x \;
\left ( {\textstyle \sqrt{g(x)} \displaystyle} \right )^{\sigma}
\; \prod_{ \mu \ge \nu } \, d g_{ \mu \nu } (x) \;
\exp \left \{ 
- \int d^4 x \, \sqrt g \, \Bigl ( \lambda - { k \over 2 } \, R
+ { a \over 4 } \, R_{\mu\nu\rho\sigma} R^{\mu\nu\rho\sigma}
+ \cdots \Bigr ) \right \}  \;\; ,
\label{eq:zcont}
\eeq
with $k^{-1} = 8 \pi G $, and $G$ Newton's constant, and reduces
to it for smooth enough field configurations.
In the discrete case the integration over metrics is replaced by
integrals over the elementary lattice degrees of freedom,
the squared edge lengths. The discrete gravitational measure in $Z_L$
can be considered as the lattice analog of the DeWitt \cite{dewitt}
continuum functional measure \cite{det}.
The $\delta A$ term in the lattice action is the well-known Regge
term \cite{regge}, and reduces to the Einstein-Hilbert action
in the lattice continuum limit \cite{rowi,hw3d}.
A cosmological constant term is needed for convergence of the path
integral, while the curvature squared term allows one to control the
fluctuations in the curvature \cite{lesh,hw84,monte,phases}.
In practice, and for obvious phenomenological reasons, one is only interested
in the limit when the higher derivative contributions are small compared
to the rest, $a \rightarrow 0$.
In this limit the theory depends, in the
absence of matter and after a suitable rescaling of the metric, only on
one bare parameter, the dimensionless coupling $k^2 / \lambda $.
Without loss of generality, one can therefore set the bare cosmological
constant $\lambda=1$.

Some partial information about the behavior of physical correlations
can be obtained indirectly from local invariant averages.
In \cite{lesh,phases} gravitational
observables such as the average curvature and its fluctuation were introduced.
The appropriate lattice analogs of these quantities are readily written down
by making use of the usual correspondences
$ \int d^4 x \, \sqrt{g} \to \sum_{\rm hinges \, h} V_h $ etc..
On the lattice the natural choices for invariant operators are 
\bea
\sqrt{g} \, (x) \; & \to & \; 
\sum_{{\rm hinges} \, h \supset x } \; V_h 
\nonumber \\
\sqrt{g} \, R (x) \; & \to & \; 
2 \sum_{{\rm hinges} \, h \supset x } \; \delta_h A_h
\nonumber \\
\sqrt{g} \, R_{\mu\nu\lambda\sigma} R^{\mu\nu\lambda\sigma} (x) \; & \to & \;
4 \sum_{{\rm hinges} \, h \supset x } \; ( \delta_h A_h )^2 / V_h 
\label{eq:ops}
\eea
(we have omitted here on the r.h.s. an overall numeric coefficient, which
will depend on how many hinges are actually included in the summation;
if the sum extends over all hinges within a single hypercube, then there
will be a total of 50 hinge contributions).
In this paper no higher derivative terms will be considered, and thus only
the first and second operators will be used in the following discussion.

On the lattice one prefers to define quantities in such a way that
variations in the average lattice spacing $\sqrt{<l^2>}$ are compensated by
the appropriate factor as determined from dimensional considerations.
In the case of the average curvature one defines the lattice quantity
${\cal R}$ as
\beq
{\cal R} (k) \; \equiv \; 
<l^2> { < 2 \; \sum_h \delta_h A_h > \over < \sum_h V_h > } \;\;\; ,
\label{eq:avr} 
\eeq
which in the continuum corresponds to
\beq
{\cal R} (k) \; \sim \;
{ < \int d^4 x \, \sqrt{ g } \, R(x) >
\over < \int d^4 x \, \sqrt{ g } > } \;\;\; ,
\eeq
and similarly for the curvature fluctuation,
\beq
\chi_{\cal R}  (k) \; \equiv \; 
{ < (\sum_h \delta_h A_h)^2 )^2 >
- < \sum_h \delta_h A_h >^2 \over < \sum_h V_h > } \;\;\; ,
\label{eq:chir} 
\eeq
which in the continuum corresponds to
\beq
\chi_{\cal R}  (k) \; \sim \;
{ < ( \int \sqrt{g} \, R )^2 > - < \int \sqrt{g} \, R >^2 \;\;\; .
\over < \int \sqrt{g} > }
\eeq
The latter is related to the connected curvature correlation at
zero momentum
\beq
\chi_{\cal R} \; 
\sim \; { \int d^4 x \int d^4 y < \sqrt{g(x)} R(x) \; \sqrt{g(y)} R(y) >_c
\over < \int d^4 x \sqrt{g(x)} > } \;\;\; .
\eeq
Both ${\cal R}$ and $\chi_{\cal R}$ are related to derivatives of $Z_L$
with respect to $k$,
\beq
{\cal R} (k) \, \sim \,
\frac{1}{V} \frac{\partial}{\partial k} \ln Z_L \; ,
\label{eq:avrz} 
\eeq
and 
\beq
\chi_{\cal R}  (k) \, \sim \,
\frac{1}{V} \frac{\partial^2}{\partial k^2} \ln Z_L \; .
\label{eq:chirz} 
\eeq
One can contrast the behavior of the preceding quantities,
associated strictly with the curvature, with the analogous quantities
involving the local volumes (and which correspond to the square root of
the determinant of the metric in the continuum).
Consider the average volume per site
\beq
\langle V \rangle  \; \equiv \; { 1 \over N_0 } < \sum_h V_h > \;\;\; ,
\label{eq:avv} 
\eeq
and its fluctuation defined as
\beq
\chi_V (k) \; \equiv \; 
{ < (\sum_h V_h)^2 )^2 >
- < \sum_h V_h >^2 \over < \sum_h V_h > } \;\;\; ,
\label{eq:chiv} 
\eeq
where one denotes by $V_h$ the volume associated with the hinge $h$.
In the continuum it corresponds to the expression
\beq
\chi_{V} (k) \, \sim \, 
{  < ( \int \sqrt{g} )^2 > - < \int \sqrt{g} >^2
\over < \int \sqrt{g} > } \;\;\; .
\eeq
The latter is related to the connected volume correlator at zero momentum
\beq
\chi_V \; \sim \; { \int d^4 x \int d^4 y < \sqrt{g(x)} \sqrt{g(y)} >_c
\over < \int d^4 x \sqrt{g(x)} > } \;\;\; .
\eeq
The average volume per site $<\!V\!>$ and its fluctuation
$\chi_V$ are simply related to derivatives of $Z_L$ with respect to the
bare cosmological constant $\lambda$,
\beq
<\!V\!> \; \sim \; \frac{\partial}{\partial \lambda} \ln Z_L \;\;\; ,
\label{eq:avvz} 
\eeq
and 
\beq
\chi_V (k) \; \sim \; \frac{\partial^2}{\partial \lambda^2} \ln Z_L \;\;\; .
\label{eq:chivz} 
\eeq
One would expect the fluctuations in the curvature to be sensitive
to the presence of a spin two massless particle,
while fluctuations in the volume would only probe the correlations in the
conformal mode channel.

\vskip 30pt
\section{Sum Rules}
\hspace*{\parindent}

In this section some useful sum rules will be derived, which follow
from simple scaling properties of the discrete functional integral.
These will be later used in the discussion of the numerical results.
A simple scaling argument, based on neglecting the effects
of curvature terms entirely (which vanish in the vicinity of the critical
point), gives first of all an estimate of the average volume per edge
\beq
< \! V_l \! > \; \sim \; { 2 ( 1 + \sigma d ) \over \lambda d }
\; \mathrel{\mathop\sim_{ d=4, \; \sigma=0 }} \;
{ 1 \over 2 \lambda } \;\; .
\eeq
In four dimensions the numerical simulations with $\sigma =0$ agree
quite well with the above formula.

Additional exact lattice identities can be obtained by examining
the scaling properties of the action and measure.
The bare couplings $k$ and $\lambda$ in the gravitational action are
dimensionful in four dimensions, but one can define the dimensionless
ratio $k^2 / \lambda$, and rescale the edge lengths so as to eliminate
the overall length scale $\sqrt{ k/\lambda} $.
As a consequence the path integral for pure gravity,
\beq
Z_L ( \lambda, k, a) \; = \; 
\int d \mu [ l^2 ] \; e^{ - I[l^2] } ,
\eeq
obeys the simple scaling property
\beq
Z_L ( \lambda, k, a) = 
\left ( { k \over \lambda } \right )^{N_1}
Z_L \left ( { k^2 \over \lambda }, { k^2 \over \lambda }, a \right ) 
\; = \;
\left ( \lambda \right )^{-N_1 / 2}
Z_L \left ( 1 , { k \over \sqrt{\lambda} }, a \right ) 
\; = \;
\left ( \lambda \right )^{-N_1}
Z_L \left ( { \lambda \over k^2 }, 1, a \right ) ,
\label{eq:scale} 
\eeq
where $N_1$ represents the number of edges in the lattice, and
the $dl^2$ measure ($\sigma=0$) has been selected \cite{det}, which is
the lattice analog of the continuum DeWitt functional measure.
This equation implies in turn a sum rule for local averages, which
(again for the specific case of the $dl^2$ measure) reads
\beq
2 \lambda < \sum_h V_h > \; - \; 
k < \sum_h \delta_h A_h > \; - \;  N_1  \; = \; 0 \;\;\; ,
\label{eq:sumr1} 
\eeq
and is easily derived from Eq.~(\ref{eq:scale}) and the definitions in
Eqs.~(\ref{eq:avrz}) and (\ref{eq:avvz}).
$N_0$ represents the number of sites in the lattice, and the
averages are defined per site (for the hypercubic lattice used in this
paper, $N_1 = 15 N_0$, $N_2 = 50 N_0$, $N_3 = 36 N_0$ and $N_4 = 24 N_0$).
The coefficients on the l.h.s. of the equation reflect the scaling
dimensions of the various terms, with the last term on the l.h.s. term
arising from the scaling property of the functional measure.
This last formula is very useful in checking the accuracy of
numerical calculations and the convergence properties of the Monte Carlo
sampling, and is usually satisfied to high accuracy $\tilde {\cal O}(10^{-4})$.
It is easy to see that a similar sum rule holds for the fluctuations,
\beq
4 \lambda^2 \left [ < (\sum_h V_h)^2 > - < \sum_h V_h >^2 \right ] \; - \; 
k^2 \left [ < (\sum_h \delta_h A_h)^2 > -  < \sum_h \delta_h A_h >^2 \right ]
\; - \; 2 N_1 \; = \; 0 \;\; .
\label{eq:sumr2} 
\eeq
Further sum rules can be derived by considering even higher derivatives
of $\ln Z_L$ with respect of $\lambda$ and $k$.
The last equation relates the fluctuation in the curvature to fluctuations
in the volumes, and thus implies a relationship between their singular parts
as well.
In particular, a divergence in the curvature fluctuation implies
a divergence of the same nature in the volume fluctuation.
In light of the previous discussion, from now on we shall consider without
loss of generality only the case of bare coupling $\lambda=1$.
As a consequence, all lengths will be tacitly expressed in units of the
fundamental microscopic length scale $\lambda^{-1/4}$.

\vskip 30pt
\section{Invariant Correlations}
\hspace*{\parindent}

In quantized gravity complications arise due to the fact that 
the physical distance between any two points $x$ and $y$ in a fixed
background geometry,
\beq
d(x,y \, \vert \, g) \; = \; \min_{\xi} \; \int_{\tau(x)}^{\tau(y)} d \tau 
\sqrt{ \textstyle g_{\mu\nu} ( \xi )
{d \xi^{\mu} \over d \tau} {d \xi^{\nu} \over d \tau} \displaystyle } \;\; ,
\eeq
is a fluctuating quantity dependent on the choice of background metric.
In addition, the Lorentz group used to classify spin states is
meaningful only as 
a local concept. Since the simplicial formulation is completely
coordinate independent, the introduction of the local Lorentz group
requires the definition of a tetrad within each simplex, and the notion
of a spin connection to describe the parallel transport of tensors
between flat simplices. Some of these aspects have recently been discussed 
from a continuum point of view in \cite{isham,modacorr,modapot}.

If the deficit angles are averaged over a number of contiguous hinges
which share a common vertex, one is naturally lead to the
connected correlator
\beq
G_R (d) \; \equiv \; < \sum_{ h \supset x } \delta_h A_h \;
\sum_{ h' \supset y } \delta_{h'} A_{h'} \;
\delta ( | x - y | -d ) >_c \; ,
\eeq
which probes correlations in the scalar curvatures
\beq
G_R (d) \; \sim \; < \sqrt{g} \; R(x) \; \sqrt{g} \; R(y) \;
\delta ( | x - y | -d ) >_c \; .
\eeq
Similarly one can construct the connected correlator
\beq
G_V (d) \; \equiv \; < \sum_{ h \supset x } V_h \;
\sum_{ h' \supset y } V_{h'} \;
\delta ( | x - y | -d ) >_c \; ,
\eeq
which probes correlations in the volume elements
\beq
G_V \; \sim \; < \sqrt{g} (x) \; \sqrt{g} (y) \;
\delta ( | x - y | -d ) >_c \; .
\eeq
The correlation length $\xi$ is defined through the
long-distance decay of the connected, invariant correlations at
fixed geodesic distance $d$.
For the curvature correlation one has, at large distances,
\beq
< \sqrt{g} \; R(x) \; \sqrt{g} \; R(y) \; \delta ( | x - y | -d ) >_c \;
\mathrel{\mathop\sim_{d \; \gg \; \xi }} \;\;
e^{-d / \xi } \;\;\;\; .
\label{eq:exp}
\eeq
At shorter distances one expects a slower, power law decay
\beq
< \sqrt{g} \; R(x) \; \sqrt{g} \; R(y) \; \delta ( | x - y | -d ) >_c \;
\mathrel{\mathop\sim_{d \; \ll \; \xi }} \;\; 
\left ( { 1 \over d } \right )^{2 n}  \;\;\;\; ,
\label{eq:pow1}
\eeq
with a power characterized by the exponent $n$.
In both cases, the distances considered are much larger than the
lattice spacing, $ d, \xi \gg l_0 $.
From scaling considerations one can show (see below) $n=4-1/\nu$.

Simple scaling arguments allow one to determine the scaling
behavior of correlation functions from the critical exponents which
characterize the singular behavior of local averages in the 
vicinity of the critical point.
A divergence of the correlation length $\xi$
\beq
\xi (k) \; \equiv \; m(k)^{-1} \;
\mathrel{\mathop\sim_{ k \rightarrow k_c}} \; A_\xi \;
( k_c - k ) ^{ -\nu }
\label{eq:mk}
\eeq
signals the presence of a phase transition, and leads to the appearance
of a singularity in the free energy $F(k)$.
The presence of a phase transition usually inferred from non-analytic terms
in invariant averages, such as the average curvature.
The curvature critical exponent $\delta$ is introduced via
\beq
{\cal R} (k) \; \mathrel{\mathop\sim_{ k \rightarrow k_c}}
- A_{\cal R} \, ( k_c - k )^\delta \;\;\;\; .
\label{eq:rsing}
\eeq
An additive constant could be added, but the evidence up to now
points to this constant being zero.
Similarly one sets for the curvature fluctuation
\beq
\chi_{\cal R} (k) \; \mathrel{\mathop\sim_{ k \rightarrow k_c}} \;
- A_{\cal R} \; ( k_c - k ) ^{ -(1-\delta) } \;\;\;\; .
\label{eq:chising}
\eeq
Scaling (Eqs.~\ref{eq:nu} ) relates the exponent $\delta$ to $\nu$,
\beq
\nu \; = \; { 1 + \delta \over d } \;\;\;\; .
\label{eq:nug}
\eeq
From such averages one can determine the value for $\nu$,
the correlation length exponent,
An equivalent result, relating the quantum expectation value
of the curvature to the physical correlation length $\xi$ , is
\beq
{\cal R} ( \xi ) \; \mathrel{\mathop\sim_{ k \rightarrow k_c}} \;
\xi^{ 1 / \nu - 4 }
\;\; ,
\label{eq:rm}
\eeq
which is obtained from Eqs.~(\ref{eq:mk}) and (\ref{eq:rsing}) using
\ref{eq:nug}).
Matching of dimensionalities in these equations is restored by
supplying appropriate powers of the Planck length $l_P=\sqrt{G}$.

It is then easy to relate the critical exponent $\nu$ to the 
scaling behavior of correlations at large distances.
The curvature fluctuation is related to the connected scalar curvature
correlator at zero momentum
\beq
\chi_{\cal R} (k) 
\sim { \int d^4 x \int d^4 y < \sqrt{g} R (x) \sqrt{g} R (y) >_c
\over < \int d^4 x \sqrt{g} > } \,
\mathrel{\mathop\sim_{ k \rightarrow k_c}} \, ( k_c - k )^{\delta -1} \; .
\eeq
A divergence in the fluctuation is then indicative of long range
correlations, corresponding to the presence of a massless particle.
Very close to the critical point one would expect for large separations
a power law decay in the geodesic distance,
\beq
< \sqrt{g} R (x) \sqrt{g} R (y) > \;
\mathrel{\mathop\sim_{ \vert x - y \vert \rightarrow \infty}} \;
\frac{1}{ \vert x-y \vert^{2n} } \;\;\;\; ,
\label{eq:rcorr}
\eeq
with the power $n$ related to the exponent $\delta$ via
$ n = \delta d /( 1 + \delta ) = d - 1 / \nu  $.
A priori one cannot exclude to possibility that some states
acquire a mass away from the critical point, in which case
one would expect the following behavior for the correlation functions,
\beq
< \sqrt{g} R (x) \sqrt{g} R (y) > \;
\mathrel{\mathop\sim_{ \vert x - y \vert \gg \xi }} \;
\exp ( - \vert x-y \vert / \xi )  \;\;\; ,
\label{eq:xio}
\eeq
where $\xi$ is the fundamental correlation length, and $m = 1/\xi$ the
associated mass.
The above equation can in fact be considered as a definition for what
is meant by the correlation length $\xi$.

\vskip 30pt
\section{Beta function and Continuum Limit}
\hspace*{\parindent}

The long distance behavior of quantum field theories is
determined by scaling behavior of the coupling constant
under a change in the momentum scale. Asymptotically
free theories such as QCD lead to vanishing gauge
couplings at short distances, while the opposite is
true for QED. In general the fixed point(s) of the renormalization
group need not be at zero coupling, but can be located at
some finite $G_c$, leading to a non-trivial fixed point
or limit cycle \cite{wilson,gross,zinn}.

In the $2 + \epsilon$ perturbative expansion for gravity
\cite{epsilon}
one analytically continues in the spacetime dimension by using
dimensional regularization, and applies perturbation theory about
$d=2$, where Newton's constant is dimensionless.
A similar method is quite successful in determining the critical properties
of the $O(n)$-symmetric non-linear sigma model above two dimensions
\cite{sigmamodel}.
In this expansion the dimensionful bare coupling is written as
$G_0 = \Lambda^{2-d} G $, where $\Lambda$
is an ultraviolet cutoff (corresponding on the lattice to a momentum
cutoff of the order of the inverse average
lattice spacing, $\Lambda \sim 1/ l_0 $).
There seem to be some technical difficulties with this expansion
due to the presence of kinematic singularities for the
graviton propagator in two dimension (the Einstein action is
a topological invariant in $d=2$), but which seem to have been
overcome recently.
A double expansion in $G$ and $\epsilon= d-2$
then leads in lowest order to
a nontrivial fixed point in $G$ above two dimensions
\beq
\beta (G) \, \equiv \, { \partial G \over \partial \log \Lambda } \, = \,
(d-2) \, G \, - \, \beta_0 \, G^2 \, + \cdots \;\; ,
\label{eq:beta} 
\eeq
with $\beta_0 > 0 $ for pure gravity.
To lowest order the ultraviolet fixed point is then at
$G_c \, = \, 1 / \beta_0 (d-2) $.
Integrating Eq.~(\ref{eq:beta}) close to the non-trivial fixed point
one obtains for $G > G_c $
\beq
m \, = \, \Lambda \, \exp \left ( { - \int^G \, {d G' \over \beta (G') } }
\right )
\, \mathrel{\mathop\sim_{G \rightarrow G_c }} \,
\Lambda \, | \, G - G_c |^{ - 1 / \beta ' (G_c) } \;\;\;\; ,
\eeq
where $m$ is an arbitrary integration constant, with the dimensions
of a mass, and which should be associated with some physical scale.
It would appear natural here to identify it with the inverse of
the gravitational correlation length ($\xi=m^{-1}$),
or some scale  associated with the average curvature.
The derivative of the beta function at the fixed point defines
the critical exponent $\nu$, which to this order is independent of $\beta_0$,
$\beta ' (G_c) \, = \, - (d-2) \, = \, - 1/ \nu $.

The previous results illustrate how the lattice continuum limit should
be taken.
It corresponds to $\Lambda \rightarrow \infty$,
$G \rightarrow G_c$ with $m$ held constant; for fixed lattice
cutoff the continuum limit is approached by tuning $G$ to $G_c$.
In four dimensions the exponent $\nu$ is defined by
\beq
m \, \mathrel{\mathop\sim_{G \rightarrow G_c }} \,
C \, \Lambda \, | \, G - G_c |^{ \nu } \;\; ,
\label{eq:massk} 
\eeq
where $m$ is proportional to the graviton mass, and $C$ is a calculable
numerical coefficient.
The value of $\nu$ determines the
running of the effective coupling $G(\mu)$, where $\mu$ is an
arbitrary momentum scale. The renormalization group tells us
that in general the effective coupling will
grow or decrease with length scale $ r = 1/ \mu$, depending on whether
$G > G_c$ or $G < G_c$, respectively.
The physical mass parameter $m$ is itself scale independent, and
obeys the Callan-Symanzik renormalization group equation
\beq
\mu \; { \partial \over \partial \mu } \; m \; = \;
\mu \; { \partial \over \partial \mu } \;
\left \{ \; C  \, \mu \, | \, G (\mu) - G_c |^{ \nu } \; \right \}
\, = \, 0  \;\;\;\; .
\label{eq:callan} 
\eeq
As a consequence,
for $G > G_c$, corresponding to the smooth phase, one expects for the
running, effective gravitational coupling \cite{phases,det}
\beq
G(r) \; = \; G(0) \left [ \; 1 \, + \, c \, ( r / \xi )^{1 / \nu} \, 
+ \, O (( r / \xi )^{2 / \nu} ) \; \right ] \;\; ,
\label{eq:running} 
\eeq
with $c$ a calculable numerical constant.
The physical mass $m = \xi^{-1}$ determines the magnitude of scaling
corrections, and plays a role similar to $\Lambda_{\overline{MS}}$ in QCD.
It cannot be determined perturbatively as it appears as an integration
constant.
Physically it separates the short distance, ultraviolet regime 
with characteristic momentum scale $ \mu$,
\beq
l_0^{-1} \gg \mu \gg m \; \;\; ,
\eeq
from the large distance, infrared region
\beq
m \gg \mu \gg L^{-1} \;\; ,
\eeq
where $L = <\!V\!>^{1/4} $ is the linear size of the system.

The exponent $\nu$ is simply related to the derivative of the beta
function for $G$ in the vicinity of the ultraviolet fixed point,
\beq
\beta ' (G_c) \, = \, - 1/ \nu \;\; .
\eeq
Thus computing $\nu$ is equivalent to computing the derivative
of the beta function in the vicinity of the ultraviolet fixed
point.
There are indications from the lattice theory that only the smooth
phase with $G>G_c$ exists (in the sense that spacetime collapses
onto itself for $G<G_c$), which would suggest that the 
gravitational coupling can only {\it increase} with distance.

One should also perhaps recall here the fact that a bare  cosmological
constant $\lambda$, which could appear in the original
action (as indicated in Eq.~(\ref{eq:zlatt})) has been scaled out, when
it was set equal to one by rescaling all the edge lengths.
If one puts it back in, then the effective Newton's constant would have to
be multiplied by that bare scale.
As a result  one obtains for the running of Newton's constant,
valid for ``short'' distances $ \mu \gg m $,
\beq
G ( r ) \, \mathrel{\mathop\sim_{ r \ll \xi }} \,
l_0^{2} \; \lambda^{-1/2}
\left [ G_c + \left ( { r \over C \xi } \right )^{ 1 / \nu} \;
+ \; \cdots \; \right ] \;\; ,
\label{eq:gmu}
\eeq
where $G_c$ is a pure number of order one, and below it will be argued that
$ 1 / \nu = 3 $.
The quantity $l_0$ is the average lattice spacing, and the correct dimensions
for $G(\mu)$ (length squared) have been restored. In addition
a bare cosmological constant $\lambda$ was re-introduced, which
was previously set equal to one in 
Eq.~(\ref{eq:zlatt}) (it fixes the overall length scale in the functional
integral over edge lengths).

\vskip 30pt
\section{Numerical Results}
\hspace*{\parindent}

Next we come to a discussion of the numerical methods employed in
this work and the analysis of the results.
As in previous work, the edge lengths are updated by a
straightforward Monte Carlo algorithm,
generating eventually an ensemble of configurations distributed
according to the action and measure of Eq.~(\ref{eq:zlatt}).
Further details of the method as applied to pure gravity
are discussed in \cite{lesh,phases}, and will not be repeated here.

In this work lattices of size
$4 \times 4 \times 4 \times 4 $
(with 256 sites, 3840 edges, 6144 simplices)
$8 \times 8 \times 8 \times 8 $
(with 4096 sites, 6144 edges, 98304 simplices)
$16 \times 16 \times 16 \times 16 $
(with 65536 sites, 983040 edges, 1572864 simplices)
were considered.
Even though these lattices are not very large, one should keep in mind
that due to the simplicial nature of the lattice there are many edges
per hypercube with many interaction terms, and as a consequence the
statistical fluctuations can be comparatively small, unless measurements
are taken very close to a critical point, and at rather large separation
in the case of the potential. The results presented here are
still preliminary, and in the future it should be possible to
repeat such calculations with improved accuracy on much larger lattices.

The topology is restricted to a four-torus (periodic
boundary conditions). We have argued before that
one could perform similar calculations with lattices employing different
boundary conditions or topology, but the universal infrared scaling
properties of
the theory should be determined only by short-distance renormalization
effects.
 
It seems reasonable that based on physical considerations one needs
to impose the constraint
that the scale of the curvature be much smaller than
the average lattice spacing, but still much larger than the overall
size of the system. In other words
\beq
< \! l^2 \! >  \;\;\; \ll \;\;\;
< \! l^2 \! > | {\cal R} |^{-1} \;\;\; \ll \;\;\; <\!V\!>^{1/2} \;\; .
\eeq
Or, that in momentum space the physical scales
should be much smaller that the ultraviolet cutoff, but much larger
than the infrared cutoff.
An equivalent requirement is then
\beq
L^{-1} \;\; \lsim \;\; m \;\; \lsim \;\; l_0^{-1} \;\; ,
\label{eq:window}
\eeq
where $L$ is the linear size of the system, $m = 1/ \xi$,
and $l_0$ the lattice spacing.
It should be kept in mind that in this model, and contrary
to ordinary gauge theories on a lattice, the lattice spacing
is a dynamical quantity.
Even close to the critical point where the curvature vanishes the lattice
is by no means regular, and the quantity $l_0 = \sqrt{<\!l^2\!>}$ only
represents an ``average'' cutoff parameter.

The bare cosmological constant $\lambda$ appearing in the
gravitational action
of Eq.~(\ref{eq:zlatt}) was fixed at $1$ (since this coupling sets
the overall length scale in the problem),
and the higher derivative coupling $a$ was set
to $0$ (pure Regge-Einstein action).
For the measure in Eq.~(\ref{eq:zlatt}) this
choice of parameters leads
to a well behaved ground state for $k < k_c \approx 0.053 $ for $a=0$
\cite{phases,monte}.
The system then resides in the `smooth' phase, with a fractal dimension
close to four; on the other hand
for $k > k_c$ the curvature becomes very large (`rough' phase),
and the lattice tends to collapse into degenerate configurations
with very long, elongated simplices \cite{lesh,hw84,monte,phases}.
For $a=0$ we investigated 22 values of $k$.

On the $16^4$ lattice 36,000 consecutive configurations were
generated for each value of $k$, and 22 different values
for $k$ were chosen. 
The results for different values of $k$ can be considered
as completely statistically uncorrelated, since
they originated from unrelated configurations.
On the smaller $8^4$ lattice 100,000 consecutive configurations were
generated for each value of $k$.
On the $4^4$ lattice 500,000 consecutive configurations were
generated for each value of $k$.
To accumulate the results, the machine ran continuously for about 14 months.


The results obtained for the average curvature ${\cal R}$
(defined in Eq.~\ref{eq:avr}) as a function
of the bare coupling $k$ are shown in Fig.\ 1.,  on lattices of
increasing size with  $4^4$, $8^4$ and $16^4$ sites.
Fig.\ 2. shows the $16^4$ data by itself.
The errors in are quite small, of the order of
a tenth of a percent or less, and are therefore not visible in the graph.

In \cite{phases} it was found that as $k$ is varied, the curvature is
negative for 
sufficiently small $k$ ('smooth' phase), and appears to go to zero 
continuously at some finite value $k_c$.
For $k \ge k_c$ the curvature becomes very large, and
the simplices tend to collapse into degenerate configurations
with very small volumes ($ <\!V\!> / <\!l^2\!>^2 \sim 0$).
This 'rough' or 'collapsed' phase 
is the region of the usual weak field expansion ($G \rightarrow 0$);
in the continuum it is characterized by the unbounded fluctuations
in the conformal mode. But there appears to be more structure to
the data.

Accurate and reproducible curvature data can only be obtained for $k$
below the instability point $k_u$ since, as already pointed
out in \cite{phases}, for $k > k_u \approx 0.053$
an instability develops, presumably associated with the unbounded
conformal mode. Its signature is typical of a sharp first order transition,
beyond which the system tunnels into the rough, elongated phase which
is two-dimensional in nature and has no physically acceptable continuum limit.
The instability is caused by the appearance of one or more localized singular
configuration, with a spike-like curvature singularity. 
It is not associated with any sort of coherent effect or the
appearance of long-range order, and remains localized around a few
lattice points. In other words, the correlation length $\xi$ remains
finite at $k_u$. At strong coupling such singular configurations
are suppressed by a lack of phase space due to the functional measure,
which imposes non-trivial constraints due to the triangle inequalities and
their higher dimensional analogs. In other language, the measure
regulates the conformal instability at sufficiently strong coupling.

It is characteristic of first order transitions that the free energy
develops only a delta-function singularity at $k_u$, with the metastable
branch developing no non-analytic contribution at $k_u$. Indeed it
is well known from the theory of first order transitions
that tunneling effects will lead to a purely imaginary
contribution to the free energy, with an essential singularity for
$k > k_u$ \cite{pbook}.
In the following we shall clearly distinguish the instability
point $k_u$ from the true critical point $k_c$. 
 
As a consequence, the non-analytic behavior of the free energy (and its
derivatives which include for example the average curvature) has to be
obtained
by {\it analytic continuation} of the Euclidean theory into the metastable
branch. This procedure, while unusual, is formally equivalent
to the construction of the continuum theory exclusively from
its strong coupling (small $k$, large $G$) expansion
\beq
Z_L (k) \; = \; \sum_{n=0}^{\infty} a_n k^n \;\; ,
\eeq
\beq
{\cal R} (k) \; = \; \sum_{n=0}^{\infty} b_n k^n \;\; .
\label{eq:series}
\eeq
Given a large enough
number of terms in this expansion, the nonanalytic behavior in the vicinity
of the true critical point at $k_c$ can then be determined using differential
or Pade approximants \cite{pade}, for appropriate combinations of thermodynamic
functions which are expected to be meromorphic in the vicinity of the true
critical point \cite{dombgreen}.
In the present case, instead of the analytic strong coupling expansion,
one has at one's disposal a set of (in principle, arbitrarily) accurate
data points to which the expected functional form can equally be fitted.
And what is assumed is the kind of regularity which is always 
assumed in extrapolating finite series (whether convergent or asymptotic
as in the case of QED or $\lambda \phi^4$ in $d<4$ \cite{qftexp})
to the boundary of their radius of convergence.

Ultimately it should be kept in mind that one is really only
interested in the {\it pseudo-Riemannian} case, and not the Euclidean
one for which an instability due to the conformal mode is
to be expected. Indeed had such an instability not occurred one
might wonder if the resulting theory still had any relationship to
the original continuum theory.  Arguments based on
effective actions suggest that if the Euclidean (or more
appropriately, Riemannian) lattice theory eventually approaches
the classical continuum theory at large distances and in the
vicinity of the critical point, then an instability in the
quantum lattice theory {\it must} develop,
since the continuum classical theory is known to be unstable.  

In the following only data for $k \le k_u$ will be considered;
in fact to add a margin of safety only $k \le 0.051$ will be
considered throughout the rest of the paper. This choice will
avoid the inclusion in the fits of any data affected by the sharp
turnover which appears, for large lattices, at $k=k_u \approx 0.053$.  

To extract the critical exponent $\delta$, one fits the computed values
for the average curvature to the form (see Eq.~\ref{eq:rsing})
\beq
{\cal R} (k) \; \mathrel{\mathop\sim_{ k \rightarrow k_c}}
- A_{\cal R} \, ( k_c - k )^\delta \;\;\;\; .
\label{eq:rsing1}
\eeq
It would seem unreasonable to expect that the computed values
for ${\cal R}$ are accurately described by this function
even for small $k$. Instead the data is fitted to the above
functional form for either $k \ge 0.02 $ or $k \ge 0.03 $ and
the difference in the fit parameters can be used as one more measure
for the error. Additionally, one can include a subleading
correction
\beq
{\cal R} (k) \; \mathrel{\mathop\sim_{ k \rightarrow k_c}}
- A_{\cal R} \, \left [ \;  k_c - k + B \; (k_c-k)^2 \; \right ]^\delta
\;\;\;\; ,
\eeq
and use the results to further constraint the errors on
$A_{\cal R}$, $k_c$ and $\delta = 4 \nu -1 $.

Using this set of procedures one obtains on the lattice with $4^4$ sites
\beq 
k_c = 0.0676(20) \;\;\;\;\;  \nu = 0.343(8)
\eeq
and on the lattice with $8^4$ sites one finds
\beq
k_c = 0.0614(27) \;\;\;\;\;  \nu = 0.322(16)
\eeq
while on the lattice with $16^4$ sites one finds
\beq
k_c = 0.0630(11) \;\;\;\;\;  \nu = 0.330(6) \;\;\;\; .
\eeq
These results suggest that $\nu$ is very close to $1/3$, and
can be compared to the older low-accuracy estimate on an $8^4$ lattice
obtained in \cite{phases} for $a=0$, $\nu=0.33(3)$.

\begin {figure}
  \begin {center}
    \input{plot1} 
  \end {center}  
\noindent
{\small Fig.\ 1 . Average curvature ${\cal R}$
as a function of $k$, on lattices with
$4^4$ ($\Box$), $8^4$ ($\triangle$) and $16^4$ ($\circ$) sites.
Statistical errors ($\sim {\cal O}(10^{-3})$)
are much smaller than the size of the symbols.
The thin-dotted, dotted and continuous lines represent best fits of the
form ${\cal R} (k) = A \; (k_c-k)^{\delta}$.
\medskip}
\end {figure}


\begin {figure}
  \begin {center}
    \input{plot2} 
  \end {center}  
\noindent
{\small Fig.\ 2 . Average curvature ${\cal R}$
as a function of $k$, on the $16^4$
($\circ$) lattice only.
Statistical errors ($\sim {\cal O}(10^{-3})$)
are much smaller than the size of the symbols.
The continuous line represents a best fit of the form $A \; (k_c-k)^{\delta}$
for $k \ge 0.02$, with $\delta = 4 \nu -1$ .
\medskip}
\end {figure}


Fig.\ 3. shows a graph of the average curvature ${\cal R}(k)$ raised to
the third power.
One would expect to get a straight line close to the critical point if
the exponent for ${\cal R}(k)$ is exactly $1/3$. The numerical data 
indeed supports this assumption, and in fact the linearity of the results
close to $k_c$ is quite striking.
The computed data is quite close to a straight line over a wide
range of $k$ values, providing further support for the assumption of
an algebraic singularity for ${\cal R}(k)$ itself, with exponent
close to $1/3$.
Using this procedure one finds on the $16^4$-site lattice 
\beq
k_c = 0.0639(10) \;\;\;\; .
\eeq


\begin {figure}
  \begin {center}
    \input{plot3} 
  \end {center}  
\noindent
{\small Fig.\ 3 . Average curvature on the $16^4$ lattice, raised to the
third power.
If $\delta=\nu=1/3$, the data should fall on a straight line.
The continuous line represents a linear fit of the form $A \; (k_c-k)$.
The small deviation from linearity of the transformed data is quite striking.
\medskip}
\end {figure}


Since the critical exponents play such a central role in
determining the existence and nature of the continuum limit,
it appears desirable to have an independent way of estimating
them, which either does not depend on any fitting procedure,
or at least analyzes a different and complementary set of data.
By studying the dependence of averages on the physical size of
the system, one can independently estimate the critical
exponents.

Fig.\ 4. shows a graph of the scaled curvature ${\cal R}(k) \; L^{4-1/\nu}$
for different values of $L=4,8,16$, versus the scaled coupling
$(k_c-k) L^{1/\nu}$. If scaling involving $k$ and $L$ holds according
to Eq.~(\ref{eq:fsso}),
with $x_{\cal O} = 1 - 4 \nu $ the scaling dimension for the curvature,
then all points should lie on the same universal curve.
From Eq.~(\ref{eq:fsso}), with $t\sim k_c-k$ and $x_O = -\delta = 1 - 4 \nu$,
one has
\beq
{\cal R} (k,L) \; = \; L^{-(4 - 1 / \nu} \; \left [ \;
\tilde{ {\cal R} } \left ( (k_c-k) \; L^{1/\nu} \right ) \; + \; 
{\cal O} ( L^{-\omega}) \; \right ]
\label{eq:fss_r}
\eeq
where $\omega>0$ is a correction-to-scaling exponent.
The data supports well such scaling behavior, and provides a further stringent
test on the value for $\nu$, which appears to be
consistent, within errors, with $1/3$.


\begin {figure}
  \begin {center}
    \input{plot4} 
  \end {center}  
\noindent
{\small Fig.\ 4 . Finite size scaling behavior of the scaled curvature
versus the scaled coupling.
Here $L=4$ for the lattice with $4^4$ sites ($\Box$), 
$L=8$ for a lattice with $8^4$ sites ($\triangle$),
and $L=16$ for the lattice with $16^4$ sites ($\circ$).
Statistical errors are comparable to the size of the dots.
The continuous line represents a best fit of the form $ a + b x^c$.
Finite size scaling predicts that all points should lie on the same universal
curve. At $k_c=0.0637$ the scaling plot gives the value $\nu=0.333$.
\medskip}
\end {figure}


Fig.\ 5. shows explicitly the size dependence of the average curvature.
For small $k$ the volume dependence is small, and gradually
increases towards the critical point. Such a trend is in agreement
with the expectation that the correlation length $\xi$ is 
growing as one approaches the critical point, leading to a more
marked volume dependence.
For fixed $k \neq k_c$ one expects on the four-torus
\beq
{\cal R}_L (k) \; \mathrel{\mathop\sim_{L \gg 1/m(k)}} \;
{\cal R}_\infty (k) \; + \; A \; m(k)^{1/2} \; L^{-3/2} \; e^{-m(k) L } 
\; + \; \cdots \;\;\; ,
\eeq
where $L=V^{1/4}$ is the linear size of the system and
$m=\xi^{-1}$ is the lightest mass in the theory.
Combining and averaging the estimates from correlations
\cite{corr}, potential \cite{phases} and finite
size corrections to the average curvature one can in fact estimate
the magnitude of this mass directly.
One obtains $ m \sim 0.81 \; (k_c-k)^{1/3}$, giving a correlation
length of about two lattice spacings at $k=0.050$.


\begin {figure}
  \begin {center}
    \input{plot5} 
  \end {center}  
\noindent
{\small Fig.\ 5. Volume dependence of the average curvature, for $L=4,8,16$,
and (from top to bottom) $k$=0.040,0.045,0.050 and 0.055.
\medskip}
\end {figure}


The value of $k_c$ itself should depend on the size of the system.
Indeed such a dependence is found when comparing $k_c$ (as obtained
from the algebraic singularity fits discussed previously) on different
lattice sizes. One writes
\beq
k_c (L) \mathrel{\mathop\sim_{L \rightarrow \infty}}
k_c (\infty) + c \; L^{-1 / \nu} \; + \; \cdots \;\;\; .
\label{eq:kcl}
\eeq
Fig.\ 6. shows the size dependence of critical coupling $k_c$ as
obtained on different size lattices.
In all three cases $k_c (L)$ is first obtained from a fit to the average
curvature of the form
${\cal R} (k) \; = \; A \; (k_c-k)^{\delta}$ for $k \ge 0.02$.
Furthermore, if one assumes $\nu=1/3$ and extracts $k_c$ from a linear fit
to ${\cal R}^3$, then the variations in $k_c$ for different size lattices
are substantially reduced (points labelled by circles in Fig.\ 6.).
Due to the few values of $L$ it is not possible at this point to extract
an estimate for $\nu$ from this particular set of data.
But since $\nu$ is close to $1/3$, it makes sense to use this value
in Eq.~(\ref{eq:kcl}) at least as a first approximation.


\begin {figure}
  \begin {center}
\setlength{\unitlength}{0.240900pt}
\ifx\plotpoint\undefined\newsavebox{\plotpoint}\fi
\sbox{\plotpoint}{\rule[-0.175pt]{0.350pt}{0.350pt}}%
\begin{picture}(1650,1320)(0,0)
\tenrm
\sbox{\plotpoint}{\rule[-0.175pt]{0.350pt}{0.350pt}}%
\put(264,158){\rule[-0.175pt]{4.818pt}{0.350pt}}
\put(242,158){\makebox(0,0)[r]{$0.06$}}
\put(1566,158){\rule[-0.175pt]{4.818pt}{0.350pt}}
\put(264,368){\rule[-0.175pt]{4.818pt}{0.350pt}}
\put(242,368){\makebox(0,0)[r]{$0.062$}}
\put(1566,368){\rule[-0.175pt]{4.818pt}{0.350pt}}
\put(264,578){\rule[-0.175pt]{4.818pt}{0.350pt}}
\put(242,578){\makebox(0,0)[r]{$0.064$}}
\put(1566,578){\rule[-0.175pt]{4.818pt}{0.350pt}}
\put(264,787){\rule[-0.175pt]{4.818pt}{0.350pt}}
\put(242,787){\makebox(0,0)[r]{$0.066$}}
\put(1566,787){\rule[-0.175pt]{4.818pt}{0.350pt}}
\put(264,997){\rule[-0.175pt]{4.818pt}{0.350pt}}
\put(242,997){\makebox(0,0)[r]{$0.068$}}
\put(1566,997){\rule[-0.175pt]{4.818pt}{0.350pt}}
\put(264,1207){\rule[-0.175pt]{4.818pt}{0.350pt}}
\put(242,1207){\makebox(0,0)[r]{$0.07$}}
\put(1566,1207){\rule[-0.175pt]{4.818pt}{0.350pt}}
\put(264,158){\rule[-0.175pt]{0.350pt}{4.818pt}}
\put(264,113){\makebox(0,0){$0$}}
\put(264,1187){\rule[-0.175pt]{0.350pt}{4.818pt}}
\put(484,158){\rule[-0.175pt]{0.350pt}{4.818pt}}
\put(484,113){\makebox(0,0){$0.05$}}
\put(484,1187){\rule[-0.175pt]{0.350pt}{4.818pt}}
\put(705,158){\rule[-0.175pt]{0.350pt}{4.818pt}}
\put(705,113){\makebox(0,0){$0.1$}}
\put(705,1187){\rule[-0.175pt]{0.350pt}{4.818pt}}
\put(925,158){\rule[-0.175pt]{0.350pt}{4.818pt}}
\put(925,113){\makebox(0,0){$0.15$}}
\put(925,1187){\rule[-0.175pt]{0.350pt}{4.818pt}}
\put(1145,158){\rule[-0.175pt]{0.350pt}{4.818pt}}
\put(1145,113){\makebox(0,0){$0.2$}}
\put(1145,1187){\rule[-0.175pt]{0.350pt}{4.818pt}}
\put(1366,158){\rule[-0.175pt]{0.350pt}{4.818pt}}
\put(1366,113){\makebox(0,0){$0.25$}}
\put(1366,1187){\rule[-0.175pt]{0.350pt}{4.818pt}}
\put(1586,158){\rule[-0.175pt]{0.350pt}{4.818pt}}
\put(1586,113){\makebox(0,0){$0.3$}}
\put(1586,1187){\rule[-0.175pt]{0.350pt}{4.818pt}}
\put(264,158){\rule[-0.175pt]{318.470pt}{0.350pt}}
\put(1586,158){\rule[-0.175pt]{0.350pt}{252.704pt}}
\put(264,1207){\rule[-0.175pt]{318.470pt}{0.350pt}}
\put(-87,682){\makebox(0,0)[l]{\shortstack{$ k_c (L) $}}}
\put(925,23){\makebox(0,0){$ 1/L $}}
\put(264,158){\rule[-0.175pt]{0.350pt}{252.704pt}}
\put(1366,515){\circle{18}}
\put(815,431){\circle{18}}
\put(539,567){\circle{18}}
\put(1366,420){\rule[-0.175pt]{0.350pt}{45.530pt}}
\put(1356,420){\rule[-0.175pt]{4.818pt}{0.350pt}}
\put(1356,609){\rule[-0.175pt]{4.818pt}{0.350pt}}
\put(815,336){\rule[-0.175pt]{0.350pt}{45.530pt}}
\put(805,336){\rule[-0.175pt]{4.818pt}{0.350pt}}
\put(805,525){\rule[-0.175pt]{4.818pt}{0.350pt}}
\put(539,504){\rule[-0.175pt]{0.350pt}{30.353pt}}
\put(529,504){\rule[-0.175pt]{4.818pt}{0.350pt}}
\put(529,630){\rule[-0.175pt]{4.818pt}{0.350pt}}
\put(1366,987){\makebox(0,0){$\triangle$}}
\put(815,546){\makebox(0,0){$\triangle$}}
\put(539,473){\makebox(0,0){$\triangle$}}
\put(1366,892){\rule[-0.175pt]{0.350pt}{45.530pt}}
\put(1356,892){\rule[-0.175pt]{4.818pt}{0.350pt}}
\put(1356,1081){\rule[-0.175pt]{4.818pt}{0.350pt}}
\put(815,452){\rule[-0.175pt]{0.350pt}{45.530pt}}
\put(805,452){\rule[-0.175pt]{4.818pt}{0.350pt}}
\put(805,641){\rule[-0.175pt]{4.818pt}{0.350pt}}
\put(539,410){\rule[-0.175pt]{0.350pt}{30.353pt}}
\put(529,410){\rule[-0.175pt]{4.818pt}{0.350pt}}
\put(529,536){\rule[-0.175pt]{4.818pt}{0.350pt}}
\sbox{\plotpoint}{\rule[-0.250pt]{0.500pt}{0.500pt}}%
\put(264,483){\usebox{\plotpoint}}
\put(264,483){\usebox{\plotpoint}}
\put(284,483){\usebox{\plotpoint}}
\put(305,483){\usebox{\plotpoint}}
\put(326,483){\usebox{\plotpoint}}
\put(347,483){\usebox{\plotpoint}}
\put(367,483){\usebox{\plotpoint}}
\put(388,483){\usebox{\plotpoint}}
\put(409,483){\usebox{\plotpoint}}
\put(430,483){\usebox{\plotpoint}}
\put(450,483){\usebox{\plotpoint}}
\put(471,483){\usebox{\plotpoint}}
\put(492,483){\usebox{\plotpoint}}
\put(513,483){\usebox{\plotpoint}}
\put(533,484){\usebox{\plotpoint}}
\put(554,484){\usebox{\plotpoint}}
\put(574,484){\usebox{\plotpoint}}
\put(595,484){\usebox{\plotpoint}}
\put(616,484){\usebox{\plotpoint}}
\put(637,484){\usebox{\plotpoint}}
\put(657,484){\usebox{\plotpoint}}
\put(678,484){\usebox{\plotpoint}}
\put(699,485){\usebox{\plotpoint}}
\put(719,485){\usebox{\plotpoint}}
\put(740,485){\usebox{\plotpoint}}
\put(761,485){\usebox{\plotpoint}}
\put(781,486){\usebox{\plotpoint}}
\put(802,486){\usebox{\plotpoint}}
\put(823,486){\usebox{\plotpoint}}
\put(843,487){\usebox{\plotpoint}}
\put(864,487){\usebox{\plotpoint}}
\put(885,487){\usebox{\plotpoint}}
\put(905,488){\usebox{\plotpoint}}
\put(926,488){\usebox{\plotpoint}}
\put(946,489){\usebox{\plotpoint}}
\put(967,489){\usebox{\plotpoint}}
\put(986,490){\usebox{\plotpoint}}
\put(1007,490){\usebox{\plotpoint}}
\put(1028,491){\usebox{\plotpoint}}
\put(1047,492){\usebox{\plotpoint}}
\put(1068,492){\usebox{\plotpoint}}
\put(1088,493){\usebox{\plotpoint}}
\put(1109,494){\usebox{\plotpoint}}
\put(1129,495){\usebox{\plotpoint}}
\put(1149,495){\usebox{\plotpoint}}
\put(1170,496){\usebox{\plotpoint}}
\put(1190,497){\usebox{\plotpoint}}
\put(1210,498){\usebox{\plotpoint}}
\put(1229,499){\usebox{\plotpoint}}
\put(1250,500){\usebox{\plotpoint}}
\put(1270,501){\usebox{\plotpoint}}
\put(1290,502){\usebox{\plotpoint}}
\put(1310,503){\usebox{\plotpoint}}
\put(1329,505){\usebox{\plotpoint}}
\put(1349,506){\usebox{\plotpoint}}
\put(1369,507){\usebox{\plotpoint}}
\put(1389,508){\usebox{\plotpoint}}
\put(1409,510){\usebox{\plotpoint}}
\put(1429,511){\usebox{\plotpoint}}
\put(1448,513){\usebox{\plotpoint}}
\put(1469,514){\usebox{\plotpoint}}
\put(1488,516){\usebox{\plotpoint}}
\put(1508,517){\usebox{\plotpoint}}
\put(1528,519){\usebox{\plotpoint}}
\put(1547,520){\usebox{\plotpoint}}
\put(1567,522){\usebox{\plotpoint}}
\put(1586,524){\usebox{\plotpoint}}
\put(264,473){\usebox{\plotpoint}}
\put(264,473){\usebox{\plotpoint}}
\put(317,473){\usebox{\plotpoint}}
\put(371,473){\usebox{\plotpoint}}
\put(425,474){\usebox{\plotpoint}}
\put(478,476){\usebox{\plotpoint}}
\put(529,480){\usebox{\plotpoint}}
\put(579,485){\usebox{\plotpoint}}
\put(629,491){\usebox{\plotpoint}}
\put(675,500){\usebox{\plotpoint}}
\put(723,510){\usebox{\plotpoint}}
\put(770,523){\usebox{\plotpoint}}
\put(816,538){\usebox{\plotpoint}}
\put(859,554){\usebox{\plotpoint}}
\put(901,572){\usebox{\plotpoint}}
\put(946,595){\usebox{\plotpoint}}
\put(986,617){\usebox{\plotpoint}}
\put(1024,642){\usebox{\plotpoint}}
\put(1061,668){\usebox{\plotpoint}}
\put(1098,696){\usebox{\plotpoint}}
\put(1134,726){\usebox{\plotpoint}}
\put(1168,757){\usebox{\plotpoint}}
\put(1200,789){\usebox{\plotpoint}}
\put(1233,823){\usebox{\plotpoint}}
\put(1265,859){\usebox{\plotpoint}}
\put(1296,896){\usebox{\plotpoint}}
\put(1327,935){\usebox{\plotpoint}}
\put(1357,975){\usebox{\plotpoint}}
\put(1385,1015){\usebox{\plotpoint}}
\put(1413,1058){\usebox{\plotpoint}}
\put(1441,1101){\usebox{\plotpoint}}
\put(1468,1144){\usebox{\plotpoint}}
\put(1493,1188){\usebox{\plotpoint}}
\put(1504,1207){\usebox{\plotpoint}}
\end{picture} 
  \end {center}  
\noindent
{\small Fig.\ 6. Volume dependence of the critical coupling $k_c$,
as determined from the singularity in the average curvature,
for lattices with $L=4,8,16$.
The points labelled by $\circ$ are obtained assuming $\nu=1/3$; in
both cases the lines represent simple fits of the type $k_c = a + b/L^3 $.
\medskip}
\end {figure}


Fig.\ 7. shows a plot of the average curvature ${\cal R}(k)$
versus reduced coupling $k_c-k$, for several values of $a$, 
the higher derivative coupling of Eq.~(\ref{eq:zlatt}).
$a=0$ corresponds to the pure Regge action with no explicit higher derivative
lattice contribution, for which
the path integral is still well defined (at least for
sufficiently small $|k|$), since the deficit
angles are bounded, and the edge lengths fluctuate around some average
value, which is determined by the interplay of the measure and the
cosmological constant term.
Alternatively, one can think of the fluctuations in the conformal
mode as becoming bounded (again at least for sufficiently small $|k|$)
when a momentum cutoff of order $\pi / \sqrt{< \! l^2 \! >}$
is dynamically generated.

The slope of each straight lines determines the critical exponent
$\delta=4 \nu-1$, and it seems clear from the graph that
the slope is noticeably smaller for $a=0$, suggesting that the
higher derivative terms mask the true critical behavior up to
very small $k_c-k$ (it was already noted in \cite{phases} that for
$a=0$ the assumption of an
algebraic singularity for the average curvature leads to a value for 
the curvature exponent which is much smaller than the estimate for
$a>0$, namely $\delta \approx 0.30(4)$).

Indeed it seems that one of the effects of the higher
derivative terms is to push the region of instability towards
smaller and smaller values of $k_c-k$, until it becomes numerically
undetectable. But we would argue that it is only close to this
region that the correct continuum behavior is recovered.
The situation is similar to what happens in the weak field
expansion and perturbation theory: higher derivative terms do
not cure the instability
problems in the physically relevant region of small momenta
and large correlation lengths.



\begin {figure}
  \begin {center}
    \input{plot8} 
  \end {center}  
\noindent
{\small Fig.\ 7. Average curvature ${\cal R}$
versus reduced coupling $k_c-k$, on a log-log scale.
From top to bottom, $a=0,0.0005,0.005,0.02,0.1$, with $a$ the
higher derivative coupling.
Statistical errobars are comparable to the size of the dots.
The slope of each straight lines determines the critical exponent
$\delta=4 \nu-1$.
The slope is noticeably smaller for $a=0$, suggesting that the
higher derivative terms mask the true critical behavior up to
very small $k_c-k$.
\medskip}
\end {figure}


Fig.\ 8. shows a plot of the curvature amplitude $A_{\cal R}$
versus the higher derivative coupling $a$. The rapid growth close
to $a=0$ is consistent with an expected catastrophic instability for $a<0$
(wrong sign for higher derivative terms).


\begin {figure}
  \begin {center}
    \input{plot9} 
  \end {center}  
\noindent
{\small Fig.\ 8.
Curvature amplitude $A_{\cal R}$ versus the higher derivative coupling
$a$. The amplitude increases rapidly as $a$ approaches zero, the pure
Einstein-Regge limit.
\medskip}
\end {figure}


A compilation of previous estimates for $\nu$, together with the
new value at $a=0$, is shown in Fig.\ 9. There seems to be a clear
trend toward smaller values as $a$ approaches zero, the Einstein-Regge 
limit.
While the Einstein action contribution becomes the dominant one
at large distances, this is no longer the case at intermediate
distances in the presence of the higher derivative terms.
One concludes that for $a > 0$ the higher derivative terms tend to mask the
true critical behavior, which requires $k_c-k \ll a^{-1}$.


\begin {figure}
  \begin {center}
\setlength{\unitlength}{0.240900pt}
\ifx\plotpoint\undefined\newsavebox{\plotpoint}\fi
\begin{picture}(1650,1320)(0,0)
\tenrm
\sbox{\plotpoint}{\rule[-0.175pt]{0.350pt}{0.350pt}}%
\put(264,158){\rule[-0.175pt]{4.818pt}{0.350pt}}
\put(242,158){\makebox(0,0)[r]{$0.3$}}
\put(1566,158){\rule[-0.175pt]{4.818pt}{0.350pt}}
\put(264,263){\rule[-0.175pt]{4.818pt}{0.350pt}}
\put(242,263){\makebox(0,0)[r]{$0.32$}}
\put(1566,263){\rule[-0.175pt]{4.818pt}{0.350pt}}
\put(264,368){\rule[-0.175pt]{4.818pt}{0.350pt}}
\put(242,368){\makebox(0,0)[r]{$0.34$}}
\put(1566,368){\rule[-0.175pt]{4.818pt}{0.350pt}}
\put(264,473){\rule[-0.175pt]{4.818pt}{0.350pt}}
\put(242,473){\makebox(0,0)[r]{$0.36$}}
\put(1566,473){\rule[-0.175pt]{4.818pt}{0.350pt}}
\put(264,578){\rule[-0.175pt]{4.818pt}{0.350pt}}
\put(242,578){\makebox(0,0)[r]{$0.38$}}
\put(1566,578){\rule[-0.175pt]{4.818pt}{0.350pt}}
\put(264,683){\rule[-0.175pt]{4.818pt}{0.350pt}}
\put(242,683){\makebox(0,0)[r]{$0.4$}}
\put(1566,683){\rule[-0.175pt]{4.818pt}{0.350pt}}
\put(264,787){\rule[-0.175pt]{4.818pt}{0.350pt}}
\put(242,787){\makebox(0,0)[r]{$0.42$}}
\put(1566,787){\rule[-0.175pt]{4.818pt}{0.350pt}}
\put(264,892){\rule[-0.175pt]{4.818pt}{0.350pt}}
\put(242,892){\makebox(0,0)[r]{$0.44$}}
\put(1566,892){\rule[-0.175pt]{4.818pt}{0.350pt}}
\put(264,997){\rule[-0.175pt]{4.818pt}{0.350pt}}
\put(242,997){\makebox(0,0)[r]{$0.46$}}
\put(1566,997){\rule[-0.175pt]{4.818pt}{0.350pt}}
\put(264,1102){\rule[-0.175pt]{4.818pt}{0.350pt}}
\put(242,1102){\makebox(0,0)[r]{$0.48$}}
\put(1566,1102){\rule[-0.175pt]{4.818pt}{0.350pt}}
\put(264,1207){\rule[-0.175pt]{4.818pt}{0.350pt}}
\put(242,1207){\makebox(0,0)[r]{$0.5$}}
\put(1566,1207){\rule[-0.175pt]{4.818pt}{0.350pt}}
\put(327,158){\rule[-0.175pt]{0.350pt}{4.818pt}}
\put(327,113){\makebox(0,0){$0$}}
\put(327,1187){\rule[-0.175pt]{0.350pt}{4.818pt}}
\put(642,158){\rule[-0.175pt]{0.350pt}{4.818pt}}
\put(642,113){\makebox(0,0){$0.05$}}
\put(642,1187){\rule[-0.175pt]{0.350pt}{4.818pt}}
\put(956,158){\rule[-0.175pt]{0.350pt}{4.818pt}}
\put(956,113){\makebox(0,0){$0.1$}}
\put(956,1187){\rule[-0.175pt]{0.350pt}{4.818pt}}
\put(1271,158){\rule[-0.175pt]{0.350pt}{4.818pt}}
\put(1271,113){\makebox(0,0){$0.15$}}
\put(1271,1187){\rule[-0.175pt]{0.350pt}{4.818pt}}
\put(1586,158){\rule[-0.175pt]{0.350pt}{4.818pt}}
\put(1586,113){\makebox(0,0){$0.2$}}
\put(1586,1187){\rule[-0.175pt]{0.350pt}{4.818pt}}
\put(264,158){\rule[-0.175pt]{318.470pt}{0.350pt}}
\put(1586,158){\rule[-0.175pt]{0.350pt}{252.704pt}}
\put(264,1207){\rule[-0.175pt]{318.470pt}{0.350pt}}
\put(-87,682){\makebox(0,0)[l]{\shortstack{ $ \nu $ }}}
\put(925,23){\makebox(0,0){ $ a $ }}
\put(264,158){\rule[-0.175pt]{0.350pt}{252.704pt}}
\put(327,310){\circle{18}}
\put(328,473){\circle{18}}
\put(330,578){\circle{18}}
\put(358,735){\circle{18}}
\put(453,735){\circle{18}}
\put(956,892){\circle{18}}
\put(327,284){\rule[-0.175pt]{0.350pt}{12.527pt}}
\put(317,284){\rule[-0.175pt]{4.818pt}{0.350pt}}
\put(317,336){\rule[-0.175pt]{4.818pt}{0.350pt}}
\put(328,420){\rule[-0.175pt]{0.350pt}{25.294pt}}
\put(318,420){\rule[-0.175pt]{4.818pt}{0.350pt}}
\put(318,525){\rule[-0.175pt]{4.818pt}{0.350pt}}
\put(330,525){\rule[-0.175pt]{0.350pt}{25.294pt}}
\put(320,525){\rule[-0.175pt]{4.818pt}{0.350pt}}
\put(320,630){\rule[-0.175pt]{4.818pt}{0.350pt}}
\put(358,630){\rule[-0.175pt]{0.350pt}{50.589pt}}
\put(348,630){\rule[-0.175pt]{4.818pt}{0.350pt}}
\put(348,840){\rule[-0.175pt]{4.818pt}{0.350pt}}
\put(453,656){\rule[-0.175pt]{0.350pt}{38.062pt}}
\put(443,656){\rule[-0.175pt]{4.818pt}{0.350pt}}
\put(443,814){\rule[-0.175pt]{4.818pt}{0.350pt}}
\put(956,814){\rule[-0.175pt]{0.350pt}{37.821pt}}
\put(946,814){\rule[-0.175pt]{4.818pt}{0.350pt}}
\put(946,971){\rule[-0.175pt]{4.818pt}{0.350pt}}
\sbox{\plotpoint}{\rule[-0.250pt]{0.500pt}{0.500pt}}%
\put(264,292){\usebox{\plotpoint}}
\put(264,292){\usebox{\plotpoint}}
\put(270,311){\usebox{\plotpoint}}
\put(275,331){\usebox{\plotpoint}}
\put(281,351){\usebox{\plotpoint}}
\put(288,371){\usebox{\plotpoint}}
\put(295,390){\usebox{\plotpoint}}
\put(302,410){\usebox{\plotpoint}}
\put(309,429){\usebox{\plotpoint}}
\put(317,449){\usebox{\plotpoint}}
\put(325,468){\usebox{\plotpoint}}
\put(333,487){\usebox{\plotpoint}}
\put(342,505){\usebox{\plotpoint}}
\put(351,524){\usebox{\plotpoint}}
\put(361,543){\usebox{\plotpoint}}
\put(370,561){\usebox{\plotpoint}}
\put(381,579){\usebox{\plotpoint}}
\put(391,597){\usebox{\plotpoint}}
\put(403,614){\usebox{\plotpoint}}
\put(415,631){\usebox{\plotpoint}}
\put(427,648){\usebox{\plotpoint}}
\put(439,664){\usebox{\plotpoint}}
\put(453,680){\usebox{\plotpoint}}
\put(466,696){\usebox{\plotpoint}}
\put(480,711){\usebox{\plotpoint}}
\put(495,725){\usebox{\plotpoint}}
\put(510,740){\usebox{\plotpoint}}
\put(525,754){\usebox{\plotpoint}}
\put(541,767){\usebox{\plotpoint}}
\put(557,781){\usebox{\plotpoint}}
\put(574,793){\usebox{\plotpoint}}
\put(591,805){\usebox{\plotpoint}}
\put(608,816){\usebox{\plotpoint}}
\put(625,827){\usebox{\plotpoint}}
\put(643,838){\usebox{\plotpoint}}
\put(661,848){\usebox{\plotpoint}}
\put(680,858){\usebox{\plotpoint}}
\put(698,867){\usebox{\plotpoint}}
\put(717,876){\usebox{\plotpoint}}
\put(736,885){\usebox{\plotpoint}}
\put(755,893){\usebox{\plotpoint}}
\put(774,901){\usebox{\plotpoint}}
\put(793,908){\usebox{\plotpoint}}
\put(813,916){\usebox{\plotpoint}}
\put(832,922){\usebox{\plotpoint}}
\put(852,929){\usebox{\plotpoint}}
\put(871,935){\usebox{\plotpoint}}
\put(891,942){\usebox{\plotpoint}}
\put(911,948){\usebox{\plotpoint}}
\put(931,953){\usebox{\plotpoint}}
\put(951,959){\usebox{\plotpoint}}
\put(971,964){\usebox{\plotpoint}}
\put(991,969){\usebox{\plotpoint}}
\put(1012,973){\usebox{\plotpoint}}
\put(1032,978){\usebox{\plotpoint}}
\put(1052,982){\usebox{\plotpoint}}
\put(1072,986){\usebox{\plotpoint}}
\put(1093,990){\usebox{\plotpoint}}
\put(1113,994){\usebox{\plotpoint}}
\put(1133,998){\usebox{\plotpoint}}
\put(1153,1002){\usebox{\plotpoint}}
\put(1174,1006){\usebox{\plotpoint}}
\put(1194,1009){\usebox{\plotpoint}}
\put(1215,1013){\usebox{\plotpoint}}
\put(1235,1016){\usebox{\plotpoint}}
\put(1256,1019){\usebox{\plotpoint}}
\put(1276,1022){\usebox{\plotpoint}}
\put(1297,1025){\usebox{\plotpoint}}
\put(1318,1027){\usebox{\plotpoint}}
\put(1338,1030){\usebox{\plotpoint}}
\put(1359,1033){\usebox{\plotpoint}}
\put(1379,1035){\usebox{\plotpoint}}
\put(1400,1039){\usebox{\plotpoint}}
\put(1420,1041){\usebox{\plotpoint}}
\put(1441,1043){\usebox{\plotpoint}}
\put(1461,1046){\usebox{\plotpoint}}
\put(1482,1048){\usebox{\plotpoint}}
\put(1503,1050){\usebox{\plotpoint}}
\put(1523,1052){\usebox{\plotpoint}}
\put(1544,1054){\usebox{\plotpoint}}
\put(1564,1056){\usebox{\plotpoint}}
\put(1585,1058){\usebox{\plotpoint}}
\put(1586,1058){\usebox{\plotpoint}}
\end{picture} 
  \end {center}  
\noindent
{\small Fig.\ 9.
Critical exponent $\nu$ computed from the average
curvature, versus the higher derivative coupling $a$.
Note the small errorbar on the recent value for $\nu$ at $a=0$.
For $a > 0$ the higher derivative terms tend to mask the
true critical behavior, which requires $k_c-k \ll a^{-1}$.
\medskip}
\end {figure}


Fig.\ 10. shows a plot of the average volume per site $<\!V\!>$, in units
of the average edge length $\sqrt{<\!l^2\!>}$.
The curve is a fit of the form $a+b(k_c-k)^c$, and suggests a
rather sudden drop of the average volume in the vicinity of the
critical point.
A non-analiticity in $<\!V\!>$ at $k_c$ is in fact consistent with the sum
rule of Eq.~(\ref{eq:sumr1}), which suggest that the singular
behavior in the average curvature ${\cal R}(k) $ and the average
volume $<\!V\!>(k)$ are simply related.
Typically, the sum rule of Eq.~(\ref{eq:sumr1}) is satisfied to
one part in $10^3$ or better.

As can be seen from Fig. 10,
close to the transition at $k_c$ the average volume per site
expressed in units of the average lattice spacing,
$<\!V\!> / <\!l^2\!>^2 $
shows only a weak singularity when the critical point is
approached from the smooth phase ($k<k_c$), and tends to a finite value.
On the other hand, in the rough phase ($k>k_c$) the volume per site
seems to approach smaller and smaller values as the lengths of the
runs are extended. In fact it would seem that in the rough phase the
volume per site can be made to approach zero, at least for some
simplices. One refers therefore alternatively to this phase
as the collapsed or polymer-like phase, since its effective
dimension is two
$^{}$\footnote{An elementary argument can be given to explain the fact
that the collapsed phase for $k>k_c$ has an effective dimension of two,
as was found in \cite{phases}. The instability is driven by the
Euclidean Einstein term in the action, and in particular its
unbounded conformal mode contribution.
As the manifold during collapse reaches an effective dimension
of two this term turns into a topological invariant, unable to drive
the instability further to a still lower dimension.}.
Furthermore the relaxation times in the rough phase become exceedingly
long, with the system
getting stuck in some degenerate, spike-like configurations without being able
to get out of it again.

It seems difficult to see how the collapse of the simplices could
be averted by choosing a different lattice structure (for example
a random lattice), since its properties seem to be unaffected
by changes in the measure or the action, at least to the extent
they have been investigated. Indeed the collapsed, polymer-like phase
appears even in the simplest models based on a regular tessellation
of the four-sphere \cite{lesh,hartle}.
From a continuum point of view,
the existence of such a pathological phase is not 
unexpected, and is interpreted as reflection of the unbounded fluctuations
in the conformal mode expected for sufficiently large $k$.
Indeed unbounded fluctuations in the conformal mode in the continuum
correspond to rapid fluctuations in the simplicial volumes, and
this is precisely what is observed on the lattice for $k>k_c$, namely a rapid
variation of simplicial volumes when going from one simplex to a
neighboring one.


\begin {figure}
  \begin {center}
    \input{plot11} 
  \end {center}  
\noindent
{\small Fig.\ 10.
Average volume per site $<\!V\!>$, in units of the average edge length.
Statistical errors are much smaller than the size of the dot.
The curve is a fit of the form $a + b\; (k_c-k)^c$. Note the resolution
on the vertical scale.
\medskip}
\end {figure}


Fig.\ 11. shows a plot of the average edge length $\sqrt{<\!l^2\!>}$.
The curve is a fit of the form $a+b(k_c-k)^c$, and suggests a
rapid increase in this quantity towards the
critical point at $k_c$. Indeed as the critical point is approached
the number of fairly small and fairly large edge lengths proliferate,
leading to an increasingly wide edge length distribution.


\begin {figure}
  \begin {center}
    \input{plot12} 
  \end {center}  
\noindent
{\small Fig.\ 11.
Average edge length as a function of the bare coupling $k$.
The curve is a fit of the form $a + b \; (k_c-k)^c$ for $k \ge 0.02$.
Statistical errors are much smaller than the symbol size.
\medskip}
\end {figure}


Fig.\ 12. shows a plot of the average curvature fluctuation
$\chi_{\cal R}(k)$ defined in Eq.~(\ref{eq:chir}).
At the critical point the curvature fluctuation diverges, by
definition.
As in the case of the average curvature 
${\cal R}(k)$ analyzed previously, one can extract the critical exponent
$\delta$ and $k_c$ by fitting the computed values
for the curvature fluctuation to the form (see Eq.~(\ref{eq:chising}))
\beq
\chi_{\cal R} (k) \; \mathrel{\mathop\sim_{ k \rightarrow k_c}}
A_{\chi_{\cal R}} \, ( k_c - k )^{-(1-\delta)} \;\;\;\; .
\label{eq:chising1}
\eeq
As for the curvature itself, 
it would seem unreasonable to expect that the computed values
for ${\cal R}$ are accurately described by this function
even for small $k$. Instead the data is fitted to the above
functional form for either $k \ge 0.02 $ or $k \ge 0.03 $ and
the difference in the fit parameters can be used as one more measure
for the error. Additionally, one can include a subleading
correction
\beq
\chi_{\cal R} (k) \; \mathrel{\mathop\sim_{ k \rightarrow k_c}}
- A_{\chi_{\cal R}}
\left [ \; k_c - k + B (k_c-k)^2 \; \right ]^{-(1-\delta)} \;\;\;\; ,
\eeq
and use the results to further constraint the errors on
$A_{\chi_{\cal R}}$, $k_c$ and $\delta = 4 \nu -1 $.

The values for $\delta$ and $k_c$ obtained in this fashion
are consistent with the ones obtained from the average curvature
${\cal R}(k)$, but with somewhat larger errors, since fluctuations
are more difficult to compute accurately than local averages, and
require much higher statistics.
Using these procedures one obtains on the lattice with $16^4$ sites
\beq
k_c = 0.0636(30) \;\;\;\;\;  \nu = 0.317(38) \;\;\;\; .
\eeq


\begin {figure}
  \begin {center}
    \input{plot13} 
  \end {center}  
\noindent
{\small Fig.\ 12. Curvature fluctuation 
on lattices with $4^4$ ($\Box$), $8^4$ ($\triangle$) and $16^4$ ($\circ$)
sites.
The thin-dotted, dotted and continuous lines represent best fits of the form
$ \chi_{\cal R} (k) \; = \; A \; (k_c-k)^{ -(1-\delta) } $ for $k \ge 0.02$.
\medskip}
\end {figure}


Fig.\ 13. shows a graph of the inverse curvature fluctuation
$\chi_{\cal R}(k)$ on the $16^4$-site lattice, raised to power $3/2$.
One would expect to get a straight line close to the critical point if
the exponent for $\chi_{\cal R}(k)$ is exactly $-2/3$. The numerical data 
indeed supports this assumption.
The computed data is consistent with linear behavior for small
$k \ge 0.02 $, providing further support for the assumption of
an algebraic singularity for $\chi_{\cal R}(k)$ itself, with exponent
close to $-2/3$.
Using this procedure one finds on the $16^4$-site lattice 
\beq
k_c = 0.0641(17) \;\;\;\; ,
\eeq
which is completely consistent with the value obtained
from ${\cal R}^{3}$ (see Fig.\ 3. and related discussion), and
suggests again that the exponent $\nu$ must be close to $1/3$.


\begin {figure}
  \begin {center}
    \input{plot14} 
  \end {center}  
\noindent
{\small Fig.\ 13. Inverse curvature fluctuation raised
to the power $3/2$, on the $16^4$ ($\circ$) lattice;
data is scaled by a factor of $\times 100$.
The straight line represents a linear fit of the form $ A \; (k_c-k) $.
The location of the critical point in $k$ is consistent with the estimate
obtained from the average curvature, but with a somewhat larger error.
\medskip}
\end {figure}


Fig.\ 14. shows the results for the logarithmic derivative
of the average curvature ${\cal R}(k)$, obtained from the data
shown in Figs.\ 3. and 12.
From the definition of the average curvature ${\cal R}$ and
curvature fluctuation (Eqs.~(\ref{eq:avr}) and (\ref{eq:chir})),
and the fact that both are proportional to derivatives of the
free energy $F$ with respect to $k$ (Eqs.~(\ref{eq:avrz}) and
(\ref{eq:chirz})),
one notices for the ratio
\beq
{ 2 \langle l^2 \rangle \; \chi_{\cal R} (k) \over {\cal R} (k) } \; \sim \;
( \frac{\partial}{\partial k} \ln Z_L ) / 
( \frac{\partial^2}{\partial k^2} \ln Z_L )
\; \sim \; \frac{\partial}{\partial k} \ln 
\left ( \frac{\partial}{\partial k} \ln Z_L \right ) \;\; .
\eeq
The assumption of an algebraic singularity in $k$ for ${\cal R}$
and $\chi_{\cal R}$ (Eqs.~(\ref{eq:rsing}) and (\ref{eq:chising}))
then implies that the logarithmic derivative as defined above
has a simple pole at $k_c$, with residue $\delta=4\nu-1$
\beq
{ 2 \langle l^2 \rangle \; \chi_{\cal R} (k) \over {\cal R} (k) }
\; \mathrel{\mathop\sim_{ k \rightarrow k_c}} \;
{\delta \over k - k_c } \;\;\;\; ,
\label{eq:pole} 
\eeq
with the critical amplitude dropping out of this particular expression.
The above result is general and does not rely on $k$ being real.
This suggests that in principle the method of Pade rational
approximants (which applies only to meromorphic functions) can be
employed to locate singularities in $\chi_{\cal R}(k) $, even for complex $k$
\cite{pade,dombgreen}.
Using this method on the $16^4$ lattice one finds
\beq
k_c = 0.0635(11) \;\;\;\;\;  \nu = 0.339(9) \;\;\;\; .
\eeq
It is encouraging that the above estimates are in good agreement
with the values obtained previously using the other methods.


\begin {figure}
  \begin {center}
    \input{plot15} 
  \end {center}  
\noindent
{\small Fig.\ 14. Inverse of the logarithmic derivative of the average
curvature ${\cal R}(k)$.
The straight line represents a best fit of the form $ A \; (k_c-k) $ for
$k \ge 0.02$.
The location of the critical point in $k$ is consistent with the estimate
coming from the average curvature ${\cal R}$.
From the slope of the line one computes directly the exponent $\nu$.
\medskip}
\end {figure}


Fig.\ 15. shows a graph of the scaled curvature fluctuation
$\chi_{\cal R}(k) / L^{2/\nu-4}$ for different values of $L=4,8,16$,
versus the scaled coupling $(k_c-k) L^{1/\nu}$.
If scaling involving $k$ and $L$ holds according to Eq.~(\ref{eq:fsso}),
with $t\sim k_c-k$ and $x_O = 1-\delta = 2 - 4 \nu$
then all points should lie on the same universal curve.
From the general Eq.~(\ref{eq:fsso}) one expects in this particular case
\beq
\chi_{\cal R} (k,L) \; = \; L^{2 / \nu - 4} \; \left [ \;
\tilde{ \chi_{\cal R} } \left ( (k_c-k) \; L^{1/\nu} \right ) \; + \; 
{\cal O} ( L^{-\omega}) \; \right ] \;\;\;\; ,
\label{eq:fss_c}
\eeq
where $\omega>0$ is again the correction-to-scaling exponent.
Again the data supports such scaling behavior, and provides
a further estimate on the value for $\nu$, close to $1/3$.


\begin {figure}
  \begin {center}
    \input{plot16} 
  \end {center}  
\noindent
{\small Fig.\ 15. Finite size scaling behavior of the scaled curvature
fluctuation versus the scaled coupling.
Here $L=4$ for the lattice with $4^4$ sites ($\Box$), 
$L=8$ for the lattice with $8^4$ sites ($\triangle$),
and $L=16$ for the lattice with $16^4$ sites ($\circ$).
The continuous line represents a best fit of the form $ 1/(a + b x^c)$.
Finite size scaling predicts that all points should lie on the same universal
curve. At $k_c=0.0637$ the scaling plot gives the value $\nu=0.318$.
\medskip}
\end {figure}


Fig.\ 16. shows a plot of the curvature fluctuation $\chi_{\cal R}$
versus the curvature ${\cal R}$. 
If the curvature approaches zero at the critical point where the
curvature fluctuation diverges, one would expect the curvature
fluctuation to diverge at ${\cal R}=0$. One has
\beq
\chi_{\cal R} ( {\cal R} )
\; \mathrel{\mathop\sim_{ k \rightarrow k_c}} \;
A \; \vert {\cal R} \vert^{(1-\delta)/\delta}
\; \sim \; A \; \vert {\cal R} \vert^{ (4 \nu -2) / (4 \nu -1) } \;\;\; .
\label{eq:cr_r}
\eeq
An advantage of this particular combination is that it does not require
the knowledge of $k_c$ in order to estimate $\nu$.
Using all points corresponding to $k \ge 0.02$ one finds
\beq
\nu = 0.328(6) \;\;\;\; .
\eeq
The error on $\nu$ can be estimated, for example, by using a more
elaborate fit of the type
\beq
\chi_{\cal R}
\; \mathrel{\mathop\sim_{ {\cal R} \rightarrow 0 }} \;
A \; \vert \; {\cal R} + B {\cal R}^2 \; \vert^{ 
(4 \nu -2) / (4 \nu -1) } \;\;\; .
\eeq
For $\nu=1/3$ the exponent becomes equal to $-2$, and one has the simple
result
\beq
\chi_{\cal R}
\; \mathrel{\mathop\sim_{ {\cal R} \rightarrow 0 }} \;
A \; \vert {\cal R} \vert^{-2} \;\;\; .
\eeq
One concludes that the evidence supports a vanishing curvature
at the critical point, where the curvature fluctuation $\chi_{\cal R}$
and the correlation length $\xi$ diverge.
This result is further supported by the consistency of the values
for $k_c$ obtained independently from ${\cal R}(k)$ and $\chi_{\cal R} (k)$
(Figs. 2,3,4,12,13,14 and 15).


\begin {figure}
  \begin {center}
    \input{plot17} 
  \end {center}  
\noindent
{\small Fig.\ 16. Inverse curvature fluctuation, $1/\chi_{\cal R}$, versus
the average curvature ${\cal R}$ ($\Box$), and $1/\sqrt{\chi_{\cal R}}$
versus 
${\cal R}$ ($\circ$). Points shown are for the largest, $16^4$-site, lattice.
For $\nu=1/3$, $1/\sqrt{\chi_{\cal R}}$ is expected to be linear in ${\cal R}$
for small ${\cal R}$.
\medskip}
\end {figure}

As an independent measure of the fluctuation one can also investigate
the behavior of the edge length fluctuation defined as
\beq
\chi_{l^2} (k) \; = \; { 1 \over N_1 } \left \{ 
\; < ( \sum_{i=1}^{N_1} l_i^2 )^2 > \; -
\; < \sum_{i=1}^{N_1} l_i^2 >^2 \right \}
\; \mathrel{\mathop\sim_{ k \rightarrow k_c}} \;
( k_c - k )^{-\gamma} \;\;\;\; ,
\label{eq:cl2}
\eeq
where $\gamma$ is a critical exponent. 
Using an analysis similar to what is done for the curvature
and curvature fluctuation, on the $16^4$ lattice it is found to diverge at
\beq
k_c = 0.0609(23) \;\;\;\;
\eeq
in agreement within errors with the previous values quoted for $k_c$.
One would expect such a fluctuation to be related to the fluctuations
in the local volumes, and, by the sum rule of Eq.~(\ref{eq:sumr1}) which
relates the fluctuations in the volume to fluctuations in the curvature,
one would expect $\gamma = 1 - \delta = 2 - 4 \nu $. The numerical
results for gamma have larger errors but give values between $0.46$
and $0.85$, certainly consistent with a value of $\gamma=2/3$ for $\nu=1/3$.


Finally Fig.\ 17. summarizes the known information about the phase diagram
in the $k$-$a$ plane. The continuous line separates the smooth phase with
small negative curvature from the rough, polymer-like phase.


\begin {figure}
  \begin {center}
    \input{plot25} 
  \end {center}  
\noindent
{\small Fig.\ 17. Phase diagram for the model in the $k-a$ plane. 
A critical line separates the smooth, strong coupling, phase from the rough,
weak coupling, phase. The dotted line denotes the pure Einstein theory,
without higher derivative terms.
\medskip}
\end {figure}


Table I summarizes the results obtained for the critical point
$k_c =1/8 \pi G_c$ and the critical exponent $\nu$.
From the best data (with the smallest statistical uncertainties
and the least systematic effects) one concludes
\beq
k_c = 0.0636(11) \;\;\;\;\;  \nu = 0.335(9) \;\;\;\; ,
\eeq
which suggests $\nu = 1/3$ for pure gravity
$^{}$\footnote{The value $\nu=1/3$ does not correspond to any
known field theory or statistical mechanics model in four
dimensions. For dilute branched polymers it is known that $\nu=1/2$
in three dimensions \cite{parsour}, and $\nu=1/4$ at the upper critical
dimension $d=8$ \cite{polymers}, so one would expect a value close to
$1/3$ somewhere in between.
I thank John Cardy for a discussion on this point.}.

\begin{table}

\begin{center}
\begin{tabular}{|l|l|l|}
\hline\hline
Method  & $k_c$ & $\nu$ 
\\ \hline \hline
${\cal R}$ vs. $k$ & 0.0630(11) & 0.330(6)
\\ \hline
${\cal R}^{3}$ vs. $k$ & 0.0639(10) & -
\\ \hline
$\chi_{\cal R}$ vs. $k$ & 0.0636(30) & 0.317(38)
\\ \hline
$\chi_{\cal R}^{3/2}$ vs. $k$ & 0.0641(17) & -
\\ \hline
$\chi_{\cal R}/(\langle l^2 \rangle {\cal R})$ vs. $k$ & 0.0635(11) & 0.339(9)
\\ \hline
$\chi_{\cal R}$ vs. ${\cal R}$ & - & 0.328(6)
\\ \hline
$\chi_{l^2}$ vs. $k$ & 0.0609(23) & $\gamma$ = 0.46(8)
\\ \hline
$\chi_{l^2}$ vs. ${\cal R}$ & - & $\gamma$ = 0.54(7)
\\ \hline \hline
${\cal R}$ FS scaling & - & 0.333(2)
\\ \hline
$\chi_{\cal R}$ FS scaling & - & 0.318(10)
\\ \hline
$\chi_{l^2}$ FS scaling & - & $\gamma$ = 0.85(6)

\\ \hline \hline
\end{tabular}
\end{center}
\label{grav}

\center{\small {\it
Table I: Summary table for the critical point $k_c$ and the critical
exponent $\nu$,
as obtained from the largest lattice with $16^4$ sites.
The last three entries assume a critical point at $k_c = 0.0636$.
\medskip}}


\end{table}
\vskip 10pt

\vskip 30pt
\section{Critical Exponents and Phenomenoloy}
\hspace*{\parindent}

In this section some consequences of the results presented above
will be discussed, with ultimately an eye towards possible physical
applications. 
Naively one would expect simply on the basis of dimensional
arguments that the curvature scale gets determined by the correlation
length
\beq
{\cal R} \; \mathrel{\mathop\sim_{ {\cal R} \rightarrow 0}} \; 1/ \xi^2 \;\; ,
\label{eq:naive}
\eeq
but one cannot in general exclude the appearance of some non-trivial exponent.

In the previous section arguments have been given in support
of the value $\nu=1/3$ for pure gravity. 
From Eqs.~(\ref{eq:rm}) relating the average curvature to the correlation
length one has
\beq
{\cal R} ( \xi ) \; \mathrel{\mathop\sim_{ k \rightarrow k_c}} \;
{ 1 \over l_P^{2-d+1/\nu} \xi^{d-1/\nu} } \;\; ,
\label{eq:rm1}
\eeq
and the correct dimension for the average curvature ${\cal R}$
have been restored by supplying
appropriate powers of the ultraviolet cutoff, the Planck length $l_P=\sqrt{G}$.
One notices that close to two dimensions the exponent of $\xi$ 
indeed approaches 2, since $\nu \sim 1/(d-2)$, and the classical result
is recovered.

For $\nu=1/3$ in four dimensions
$^{}$\footnote{For all scalar field theories (spin $s=0$) in four
dimensions it is known that $\nu=1/2$, while for the compact Abelian U(1)
gauge theory ($s=1$) one has $\nu=2/5$ \cite{u1}. The value $\nu=1/3$
for pure gravitation ($s=2$) in four dimensions is then consistent with
the simple formula $\nu = 1/(2+s/2)$.}
one then obtains the remarkably simple result
\beq
{\cal R} ( \xi ) \; \mathrel{\mathop\sim_{ k \rightarrow k_c}} \;
{ 1 \over l_P \; \xi } \;\;\;\; .
\label{eq:rm2}
\eeq
An equivalent form can be given in terms of the curvature scale $H_0$,
defined through $R=-12 H_0^2$, and which has dimensions of a mass
squared. One has close to the critical point
\beq
H_0^2 \; = \; C_H \; \mu_P \; m \;\; ,
\label{eq:hub}
\eeq
where $\mu_P = 1/\sqrt{G}$ is the Planck mass, $m=1/\xi$ is the
inverse gravitational correlation length, and $C_H \approx 4.9$ a numerical
constant of order one; the value for $C$ is extracted from the
known numerical values for ${\cal R}$ and $m$ close to the critical
point at $k_c$.

One can raise the legitimate concern of how these results are changed
by quantum fluctuations of matter fields.
In the presence of matter fields coupled to gravity (scalars, fermions,
vector bosons, spin-3/2 fields etc.) one expects the value for $\nu$
to change due to vacuum polarization loops containing these fields.
A number of arguments can be given though for why these effects
should not be too dramatic, unless the number of light matter fields is
very large.
First, in the case of a single light scalar field the vacuum polarization
effects are so small that they are barely detectable in the numerical
evaluations of the path integral \cite{scalar}.
Furthermore one notices that to leading order in the $2+\epsilon$ expansion
the exponent $\nu$ only depends on the dimensionality of space-time,
irrespective of the number of matter fields and of their type \cite{epsilon}.
Finally one can compute for example
the effects of scalar matter fields on the one-loop beta function in the
$2+\epsilon$ expansion for gravity, and finds $\beta_0 = (2/3)(25-n_f)$
where $n_f$ is the number of massless scalar fields \cite{epsilon}. 
Thus unless $n_f$ is large, the matter contribution is quite small
even to next-to-leading order in the $2+\epsilon$ expansion.
The present evidence would therefore suggest that the approximation
in which vacuum polarization effects of light matter fields are
neglected should not be too unreasonable.

It seems natural to identify $H_0$ with either some (negative) average
spatial curvature, or possibly with the Hubble constant determining the
macroscopic expansion rate of the present universe \cite{phases,det}.
In the Friedmann-Robertson-Walker model of standard cosmology \cite{weibook}
on has for the Ricci scalar
\beq
R_{Ricci} \; = \; -6 \left \{  \left ( { \dot{R} \over R } \right )^2 \; + \;
{ k \over R^2 } \; + \; { \ddot{R} \over R }  \right \} \;\;\; ,
\label{eq:ricci} 
\eeq
where $R(t)$ is the FRW scale factor, and $k=0,\pm 1$ for spatially
flat, open or closed universes respectively.
Today the Hubble constant is given by $H_0^2 = ( \dot{R} / R )_{t_0}^2 $,
but it is eventually expected to show some slow variation in time, and
its characteristic length scale $c H_0^{-1} \approx 10^{28} cm$ today
is comparable to the present extent of the visible universe.
Under such circumstances from Eq.~(\ref{eq:hub}) one would expect
the gravitational correlation $\xi$ to be significantly larger than
$c H_0^{-1}$.
A potential problem arises though in trying to establish a relationship
between quantities which are truly constants (such as the ones appearing
in Eq.~(\ref{eq:hub})), and $H_0$ which most likely depends on time.
$^{}$\footnote{The only exception being the steady state cosmological
models, where $H$ is truly a constant of nature.
These models are not favored by present observations, including
detailed features of the cosmic background radiation.}
In any case it is clear that some of these considerations
are in fact quite general, to the extent that they rely on general
principles of the renormalization group and are not tied to any
particular value of $\nu$, although $\nu=1/3$ clearly has some aesthetic
appeal.
Additional cosmological and astrophysical arguments and proposed tests
can be found in a recent paper \cite{vipul}.

One further observation can be made regarding the running of $G$.
Assuming the existence of an ultraviolet fixed point,
the effective gravitational coupling is given by Eq.~(\ref{eq:running})
for ``short distances'' $r \ll \xi$, but now with an exponent $\nu=1/3$,
\beq
G(r) \; = \; G(0) \left [ \, 1 \, + \, c \, ( r / \xi )^{3} \, 
+ \, O (( r / \xi )^{6} ) \, \right ] \;\; ,
\label{eq:run1} 
\eeq
with $c$ a calculable numerical constant of order one.
The appearance of $\xi$ in this equation, which is a very large
quantity by Eq.~(\ref{eq:hub}), suggests that the leading
scale-dependent correction, which gradually increases the strength of the
effective gravitational interaction as one goes to larger and
larger length scales, should be extremely small.
$^{}$\footnote{And suggests that the deviations from classical general
relativistic behavior for most physical quantities is in the end
practically negligible.}

It is only for distances comparable to or larger than $\xi$ that the
gravitational potential should start to weaken and fall off exponentially,
with a range given by the gravitational correlation length $\xi$,
\beq
V(r) \; \mathrel{\mathop\sim_{ r \; \gg \; \xi }} - \; G(r) \;
{ \mu_1 \mu_2 \;  e^{-r/\xi } \over r } \;\; .
\label{eq:potexp}
\eeq
In many ways these results appear qualitatively consistent with the expected
behavior of
the tree-level graviton propagator in anti-de Sitter space \cite{hawpag,ads}. 
In the real world the range $\xi$ must be of course very large.
From the fact
that super-clusters of galaxies apparently do form, one can easily set an
observational lower limit $ \xi > 10^{25} cm$.

It is unclear to what extent gravitational correlations can be 
measured directly.
From the definition of the curvature correlation function 
in Eq.~(\ref{eq:pow1}) one has for ``short distances'' $r \ll \xi$
and for the specific value $\nu=1/3$ the remarkably simple result
\beq
< \sqrt{g} \; R(x) \; \sqrt{g} \; R(y) \; \delta ( | x - y | -d ) >_c \;
\mathrel{\mathop\sim_{d \; \ll \; \xi }} \;\; 
{A \over d^2 } \;\;\; ,
\label{eq:pow2}
\eeq
with $A$ a calculable numerical constant of order one.
One can contrast this behavior with the semiclassical result attained
close to two dimensions (and which incidentally coincides with the
lowest order weak field expansion result \cite{modacorr}), which gives instead
for the power the value $ 2(d-1/\nu) \sim 2 (d - (d-2)) \sim 4 $, as
expected on the basis of naive dimensional arguments ($R \sim \partial^2 h$).

If one considers the curvature $R$ averaged over a spherical volume
$V_r = 4 \pi r^3 /3$,
\beq
\overline{ \sqrt{g} \; R } \; = \; { 1 \over V_r } \;
\int_{V_r} d^3 \vec{x} \;  \sqrt{g(\vec{x},t)} \; R(\vec{x},t)
\eeq
one can compute the corresponding variance in the curvature
\beq
\left [ \delta ( \sqrt{g} \; R ) \right ]^2 \; = \; { 1 \over V_r^2 } \;
\int_{V_r} d^3 \vec{x} \; \int_{V_r} d^3 \vec{y} \;
< \sqrt{g} \; R(\vec{x}) \; \sqrt{g} \; R(\vec{y}) >_c \; = \; 
{9 A \over 4 \; r^2 } \;\;\; .
\eeq
As a result the r.m.s. fluctuation of $\sqrt{g} R$ averaged over a 
spherical region of size $r$ is given by
\beq
\delta ( \sqrt{g} \; R ) \; = \; 
{ 3 \sqrt{A} \over 2 } \; {1 \over r } \;\;\; ,
\eeq
while the Fourier transform power spectrum at small $\vec{k}$ is 
\beq
P_{\vec{k}} \; = \; \vert \; \sqrt{g} \; R_{\vec{k}} \; \vert^2 \; = \; 
{ 4 \pi^2 A \over 2 V } \; {1 \over k } \;\;\; .
\eeq

One can use Einstein's equations to relate the local curvature to the
(primordial) mass density. From Einstein's field equations
\beq
R_{\mu\nu} - {1 \over 2} g_{\mu\nu} R \; = \; 8 \pi T_{\mu\nu} \;\;
\label{eq:ein}
\eeq
for a perfect fluid
\beq
T_{\mu\nu} \; = \; p g_{\mu\nu} + (p+\rho) u_\mu u_\nu \;\;
\label{eq:fluid}
\eeq
one obtains for the Ricci scalar, in the limit of negligible pressure,
\beq
R (x) \; \approx \; 8 \pi G \; \rho (x) \;\;\; .
\eeq
As a result one expects for the density fluctuations a power law decay
of the form
\beq
< \rho (x) \; \rho (y) >_c \;
\mathrel{\mathop\sim_{ |x-y| \; \ll \; \xi }} \;
{1 \over |x-y|^2 } \;\; .
\label{eq:pow3}
\eeq
Similar density correlations have been estimated from observational data
by analyzing known galaxy number density distributions, giving a
value for the exponent of about $1.77 \pm 0.04$ for distances in the $10kpc$
to $10Mpc$ range \cite{peebles}.

\vskip 30pt
\section{Concluding Remarks}
\hspace*{\parindent}

Numerical simulation methods combined with
modern renormalization group arguments and finite size scaling
can provide detailed information on non-perturbative aspects of
a lattice model of quantum gravity.
It has been known for some time that the lattice model has two phases,
only one of which is physically acceptable. 
In this work we have described in some detail the properties of the
latter smooth phase, and provided quantitative estimates for the critical
point, the scaling dimensions and the behavior of correlations at distances
large compared to the cutoff.
In spite of the fact that the Euclidean theory becomes unstable as one
approaches the critical point at $k_c$, it is still possible to
determine by a straightforward analytic continuation the physical properties
of the model in the vicinity of the true fixed point,
defined as the point where a non-analiticity develops in the 
strong coupling branch of $Z_L(k)$, and where scaling implies that
the physical correlation $\xi$ diverges.

If this prescription is followed, an estimate for the non-perturbative
Callan-Symanzik beta function in the vicinity of the fixed point
can be obtained, to leading order in the deviation of the bare coupling
from its critical value.
The resulting scale evolution for the gravitational constant is then
quantitatively quite small, if one assumes that the scaling violation
parameter is related to an average curvature and its characteristic
scale $H_0$.
Its infrared growth, consistent with the general idea that gravitational
vacuum polarization effects cannot exert any screening, suggests
that low energy properties of quantum gravity are inaccessible by
weak coupling perturbation theory: low energy quantum gravity
is a strongly coupled theory.
On a more quantitative side, as pointed out in the discussion there are
a number of attractive features to the pure gravity result $\nu =1/3$,
including a simple form for the curvature correlations at short distances. 

It seems legitimate to ask the question whether the present lattice model for
quantum gravity provides any insight into the problem of the cosmological
constant. The answer is both yes and no. 
To the extent that a naive prediction of quantum gravity is that
the curvature scale should be of the same order of the Planck 
length, ${\cal R} \sim 1/G$, the answer is definitely yes.
Indeed it can be regarded as a non-trivial result of the lattice models for
gravity that a region in coupling constant space can be found where
space-time is stiff and 
the curvature can be made much smaller than $1/G$. In fact the evidence
indicates that the average curvature ${\cal R}$ vanishes at the critical
point $k_c$.
And this is achieved with a bare cosmological constant $\lambda$ which
is of order one in units of the cutoff.
Phrased differently, the dimensionless ratio between the renormalized
and the bare cosmological constant becomes arbitrarily small towards
the critical point.

At the same time the effective long distance cosmological constant is
non-vanishing and of order $1/\xi$, and the value zero is only obtained
when $\xi$ is exactly zero, which happens only at the critical point $k_c$. 
Thus to make the effective cosmological constant small requires a fine
tuning, in the sense that the bare coupling $k_c-k$ has to be small.
But since the correlation length determines the corrections to the Newtonian
potential (and in particular its eventual decrease for large enough
distances), it would seem unnatural to have a short correlation
length $\xi$: in such a world there would be no long-range gravitational
forces, and separate space-time domains would have decoupled fluctuations.
From this perspective, long range forces and a small cosmological
constant go hand in hand. Quantum fluctuation effects show that
hyperbolic space-times with small curvature radii
cannot sustain long-range gravitational forces, at least in this model.

\vspace{20pt}

{\bf Acknowledgements}

The author acknowledges useful discussions with John Cardy, James Hartle,
Gary Horowitz, Ruth Williams and Vipul Periwal.
The Aeneas Supercomputer Project is supported in part by
the Department of Energy, the National Science Foundation and
the University of California.
An earlier part of the calculation described in this paper was
performed at the NSF-supported Cornell Theory Center.
Part of the work presented here was done while the author
was a visitor in the Theoretical Division at CERN. The author is grateful
to Gabriele Veneziano for discussions and hospitality.

\vfill

\newpage
\end{document}